\documentclass[a4paper]{JHEP3}
\usepackage{epsfig,bbm}
\usepackage{ifthen}
\usepackage{epsfig}
\usepackage{amssymb}
\usepackage{graphicx}
\usepackage{amsmath}
\usepackage[center]{subfigure}
\renewcommand{\text}[1]{%
\ifthenelse{\equal{#1}{mB}}{m_B}{}%
\ifthenelse{\equal{#1}{sig}}{\sigma}{}%
\ifthenelse{\equal{#1}{mc}}{m_c}{}%
\ifthenelse{\equal{#1}{q2}}{q^2}{}%
\ifthenelse{\equal{#1}{om}}{\omega}{}%
}
\newcommand{\ba}{\begin{eqnarray}}
\newcommand{\ea}{\end{eqnarray}}
\newcommand{\be}{\begin{equation}}
\newcommand{\ee}{\end{equation}}

\newcommand{\A}{{\cal A}}
\newcommand{\V}{{\cal V}}
\newcommand{\DS}[1]{/\!\!\!#1}

\title{Charm-loop effect in $B \to K^{(*)} \ell^{+} \ell^{-}$ and $B\to
K^*\gamma$}

\author{A.~Khodjamirian, Th.~Mannel, A.A. Pivovarov\footnote{\it On leave
from Institute for Nuclear Research, 117312 Moscow, Russia} \,\,
and Y.-M.~Wang \\
\it  Theoretische Physik 1, Fachbereich Physik, Universit\"at
Siegen,\\ D-57068 Siegen, Germany }

\abstract{ We calculate the long-distance effect generated by the
four-quark operators with $c$-quarks in the $B\to K^{(*)}
\ell^+\ell^-$ decays. At the lepton-pair invariant masses far
below the $\bar{c}c$-threshold, $q^2\ll 4m_c^2$, we use OPE near
the light-cone. The nonfactorizable soft-gluon emission from
$c$-quarks is cast in the form of a nonlocal effective operator.
The $B\to K^{(*)}$ matrix elements of this operator are calculated
from the QCD light-cone sum rules with the $B$-meson distribution
amplitudes. As a byproduct, we also predict the charm-loop
contribution to $B\to K^*\gamma$ beyond the local-operator
approximation. To describe the charm-loop effect at large $q^2$,
we employ the hadronic dispersion relation with $\psi=J/\psi,\psi
(2S), ...$ contributions, where the measured $ B\to K^{(*)}\psi $
amplitudes are used as inputs. Matching this relation to the
result of QCD calculation reveals a destructive interference
between the $J/\psi$ and $\psi(2S)$ contributions. The resulting
charm-loop effect is represented as a $q^2$-dependent correction
$\Delta C_9(q^2)$  to the Wilson coefficient $C_9$. Within
uncertainties of our calculation, at $q^2$ below the charmonium
region the predicted ratio $\Delta C_9(q^2)/C_9$ is $\leq 5\% $
for $B\to K \ell^+\ell^-$, but can reach as much as 20\%  for
$B\to K^*\ell^+\ell^-$, the difference being mainly caused by the
soft-gluon contribution. }


\begin{document}


\section{Introduction}

Being  very prospective channels for the search for new physics, the
exclusive $B\to K ^{(*)}\ell^+\ell^-$ decays  have quite a complicated
dynamics. In the Standard Model the underlying flavour-changing
$b\to s \ell^+\ell^-$  transition
is described by the effective Hamiltonian \cite{Heff,BBL}:
\be
H_{eff}= -\frac{4G_{F}}{\sqrt{2}}V_{tb}V_{ts}^{*}
{\sum\limits_{i=1}^{10}} C_{i}({\mu}) O_{i}({\mu})\,,
\label{eq:Heff}
\ee
a superposition of the effective operators $O_i$, weighted by their
Wilson coefficients $C_i$ and normalized at the scale $\mu\sim m_b$.
The relevant operators are presented in Appendix A.
In this paper we neglect the CKM-suppressed contributions
proportional to $V_{ub}V_{us}^*$, adopting the approximation
$|V_{tb}V_{ts}^{*}|\simeq |V_{cb}V^*_{cs}|=0.0428^{+0.001}_{-0.004}$ \cite{CKMfit}.

The decay amplitude
 \be
 A(B \to K^{(*)} \ell^+\ell^-)=
-\langle K^{(*)} \ell^+\ell^-\mid H_{eff}\mid B\rangle \,,
 \label{eq:amplA}
 \ee
contains  a rich variety of hadronic matrix elements
of the operators $O_i$.
The dominant contributions to (\ref{eq:amplA}) are generated by the
$O_{9,10}$ and $O_{7\gamma}$ with large  Wilson coefficients.
In Appendix B the hadronic matrix elements
of these operators  are presented. They are factorized in terms of the
$B\to K^{(*)} $ form factors, which, similar  to  the $B\to \pi,\rho $
form factors of the weak semileptonic decays, are
obtained from lattice QCD or QCD light-cone sum rules (LCSR).
In addition, there are specific
contributions to the $B\to K^{(*)} \ell^+\ell^-$ and $B\to K^*\gamma$
amplitudes
generated by  the current-current operators,
$O_{1,2}$ and penguin operators $O_{3-6,8g}$,
combined with the electromagnetic (e.m.) interaction of quarks.
A major challenge for the theory is to
identify and estimate these hadronic matrix elements  one by one.

An important effect, which is the main topic of our study, is
generated by the current-current operators $O_{1,2}$ acting
together with the $c$-quark e.m. current. This mechanism involves
an intermediate ``charm-loop", coupled to the lepton pair via the
virtual photon. The  analogous effects for the $u$-quark
current-current and quark-penguin operators with different
flavours are not important in $B\to K^{(*)} \ell^+\ell^-$, being
either CKM suppressed, or multiplied by a small Wilson
coefficient.

The $c$-quark loop turns into a genuine long-distance hadronic
effect if the lepton-pair invariant mass
$q^2=(p_{\ell^+}+p_{\ell^-})^2$ reaches the region of charmonium
resonances $\psi=\{J/\psi,\psi(2S), ....\}$. At $q^2=m_\psi^2$,
the process $B\to K^{(*)}\ell^+\ell^-$ transforms into a
nonleptonic weak decay $B\to\psi K^{(*)}$, followed by the
leptonic annihilation of $\psi$. All $\psi$-resonances with
$m_\psi<(m_B-m_{K^{(*)}}) $ contribute to this mechanism.
Moreover, above the $\bar{D}D$ threshold there are continuum
contributions of the intermediate charmed hadrons and it is very
difficult to include them in a model-independent way. Although the
$q^2$-intervals around $J/\psi$ and $\psi(2S)$ are subtracted from
the measured lepton-pair mass distributions in $B\to
K^{(*)}\ell^+\ell^-$, the intermediate and/or virtual $\bar{c}c$
states contribute outside the resonance region and their effect
has to be accurately estimated.

Usually, in the leading-order, the $c$-quark loop diagram is
included into the factorization formula for $B\to
K^{(*)}\ell^+\ell^-$. In addition, hard-gluon exchanges between
the $c$-quark loop and the rest of the diagram are taken into
account, together with other perturbative nonfactorizable effects
(see e.g., \cite{BFS}). One generally predicts these effects to be
small, if $q^2$ is far below the charmonium region.

Two important questions concerning
the charm-loop effect remain unanswered, despite  many
dedicated studies. The first question is: how important are
the soft gluons emitted from the $c$-quark loop
and violating the factorization?
The second question concerns the validity of the approximation
``$c$-quark-loop plus corrections'' at large $q^2$,
approaching the charmonium resonance region.
It is the purpose of our paper to address these questions.

The soft-gluon emission from the charm loop was calculated
for the inclusive $B\to X_s\gamma$  decay in
\cite{Vol} and, independently, for the exclusive channel $B\to K^*\gamma$
in \cite{KRSW}.  The main outcome was an effective
quark-antiquark-gluon operator whose Wilson coefficient is
proportional to $1/m_c^2$. In $B\to K^*\gamma $
the hadronic matrix element  of this operator was obtained
\cite{KRSW} using three-point QCD sum rules and, more recently,
LCSR \cite{BZJ}. Studies of this effect
in $B\to X_s\gamma$ and $B\to X_s\ell^+\ell^- $ \cite{cloop}
revealed that, in addition to the
lowest-dimension local operator,
a tower of operators with the derivatives
of the gluon field has to be taken into account.
Explicitly, the soft-gluon momentum
was taken into account in \cite{BIR} for the inclusive decay
$B\to X_s\ell^+\ell^-$.

In this paper we employ  operator-product expansion (OPE) near
the light-cone  for the gluon emission
from the $c$-quark loop at  $q^2\ll 4m_c^2$.
From this expansion we derive an effective nonlocal quark-antiquark-gluon
operator. Sandwiching this operator between the $B$ and $K^{(*)}$ states
we calculate the hadronic matrix
elements employing LCSR  with $B$-meson distribution amplitudes.
These sum rules have been used in \cite{KMO1,KMO2} to obtain
the $B\to $ light-meson form factors. The calculation presented below
provides an effective resummation of the soft-gluon part
of the charm-loop effect in  $B\to K^{(*)}\ell^+\ell^-$ and (as a byproduct)
in  $B\to K^* \gamma$. Adding the calculated soft-gluon contribution
to the leading-order factorizable $\bar{c}c$ loop ,
we obtain an estimate of the charm-loop effect in
$B\to K^{(*)}\ell^+\ell^-$,
valid at  $q^2\ll 4m_c^2$.

The second problem investigated in this paper is the validity of
OPE at large $q^2$, approaching the charmonium region. This issue
is closely related to the status of quark-hadron duality for the
charm-loop amplitude. Earlier models \cite{AMM}, adding  the
contributions of $\psi$-resonances on the top of the $c$-quark
loop, involve double counting of quark-gluon and hadronic degrees
of freedom. More consistent is the use of the hadronic dispersion
relation written in terms of $\psi$ states, as suggested for $B\to
X_s \ell^+\ell^-$ in \cite{KS}. Recently, this approach was
reconsidered in \cite{BBNSnew}. In all previous analyses involving
dispersion relation, nonfactorizable contributions to the
charm-loop effect were neglected, and, correspondingly, the
nonleptonic $ B\to \psi K^{(*)}$ amplitudes were taken in the
factorization approximation. In order to adjust these amplitudes
to their measured values, additional $k$-factors were introduced
\cite{KS,ABHH}.

In this paper we employ  the hadronic dispersion relation
in a different way. We fix the absolute values of residues
of the  $J/\psi$- and $\psi(2S)$-poles from the
experimental data on $B\to J/\psi K^{(*)}, \psi(2S)K^{(*)} $.
The integral over the spectral density above the open charm threshold
is parameterized in a form of an effective pole.
We then fit the whole dispersion relation to the OPE result
at $q^2\ll 4m_c^2$, including the newly calculated nonfactorizable
contribution. Importantly, we find that this procedure
favours a destructive interference between the $\psi(2S)$
and $J/\psi$ terms in the dispersion relation.
Our main result for the charm-loop effect is obtained
in terms of the $q^2$- and process-dependent correction to
the Wilson coefficient $C_9$. This correction is directly
calculated at small $q^2$ and analytically continued to
large $q^2<m_{\psi(2S)}^2$ via OPE-controlled dispersion relation.

The plan of this paper is as follows. Sect.~2 contains an
introductory discussion of the OPE for the charm-loop effect. In
Sect.~3  we expand the product of the four-quark operators with
the $c$-quark e.m. current near the light cone and obtain the
non-local operator corresponding to the nonfactorizable soft-gluon
emission. Sect.~4 contains the derivation of  LCSR for the $B\to
K$ and $B\to K^{*}$ matrix elements of this operator. Adding them
to the known factorizable matrix elements, we obtain the full
charm-loop contribution to $B\to K^{(*)} \ell^+\ell^-$, valid at
small $q^2$. In Sect.~5 we present the details of our numerical
analysis and obtain the results for the charm-loop effect in a
form of the correction to $C_9$.  Sect.~6 contains our prediction
for the charm-loop effect in $B\to K^*\gamma$. In Sect.~7   we
perform the  matching of the OPE result to the hadronic dispersion
relation in terms of charmonium resonances, predicting the
charm-loop effect at $q^2$ up to the open charm threshold.
Finally, we investigate the influence of this effect on the
observables of $B\to K^{(*)} \ell^+\ell^-$. Sect.~8 contains the
concluding discussion. In Appendix~A the effective operators and
their Wilson coefficients are collected. In Appendix~B  the
definitions of $B\to K^{(*)} \ell^+\ell^-$  amplitudes and the
inputs for form factors are presented. In Appendix~C the relevant
characteristics of $J/\psi$ , $\psi(2S)$ states and $B\to
K^{(*)}J/\psi(\psi(2S))$ decays are given. In Appendix D the
expressions for the coefficients entering the LCSR are presented.

 \section{ Light-cone dominance of the c-quark loop }

The combined action of the four-quark operators $O_1$ and $O_2$
in (\ref{eq:Heff}) and the e.m. interactions
of $c$-quarks and leptons leads to the charm-loop effect
depicted in Fig.~\ref{fig:diag}.
\FIGURE[t]{
\includegraphics[scale=0.5]{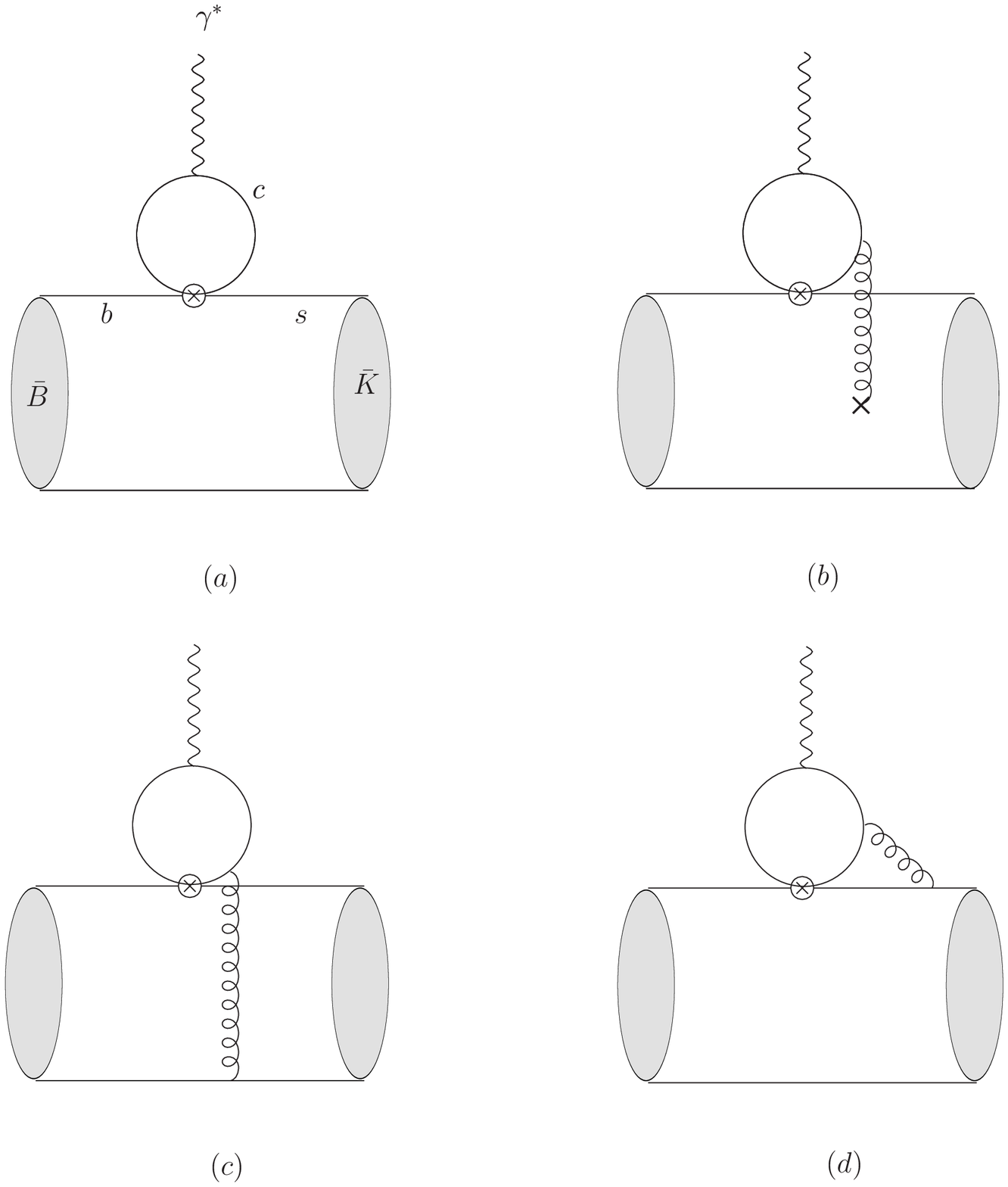}
\caption{\it Charm-loop effect in $B\to K ^{(*)}\ell^+\ell^-$:
(a)-the leading-order factorizable contribution; (b)
nonfactorizale soft-gluon emission, (c),(d)-hard gluon exchange.}
\label{fig:diag}
}
The contribution of this mechanism
to the $B \to K^{(*)} \ell^+ \ell^-$ decay amplitude can be written as
\be
A(B\to  K^{(*)} \ell^+\ell^-)^{(O_{1,2})}= -(4\pi\alpha_{em} Q_c)\frac{4G_F}{\sqrt{2}}
V_{tb}V^{*}_{ts}\frac{\bar{\ell}\gamma^\mu \ell }{q^2}
{\cal H}_\mu^{(B\to  K^{(*)})}(p,q)\,,
\label{eq:ampl}
\ee
where $Q_c=2/3$ is the $c$-quark  electric charge,  the lepton current and
photon propagator are factored out and
the hadronic transition matrix element is:
\begin{eqnarray}
{\cal H}^{(B\to  K^{(*)})}_\mu(p,q)= i\int d^4xe^{iq\cdot x} \langle  K^{(*)}(p)|
T \Big\{\bar{c}(x)\gamma_\mu c(x)\,,
\nonumber
\\
\Big[ C_1O_1(0)+C_2O_2(0)\Big] \Big\}| B(p+q) \rangle \,.
\label{eq:matr}
\end{eqnarray}

Isolating in (\ref{eq:matr}) the $T$-product of the $c$-quark e.m.
current and the $c$-quark fields entering $O_{1}$ or $O_2$, one
has in both cases a generic expression: \be {\cal
C}^a_{\mu}(q)=\int d^4xe^{iq\cdot x} T \Big\{\bar{c}(x)\gamma_\mu
c(x) ,\bar{c}_L(0)\Gamma^a c_L(0) \Big\}\,, \label{eq:Pi} \ee
where $\Gamma^a$ denotes a certain combination of Dirac- and
colour-matrices.

At $q^2\ll 4m_c^2$, that is, if $c$-quarks are
highly virtual, the dominant region of integration over $x$
in (\ref{eq:Pi})
is concentrated near the light-cone $x^2=0$. To see that, we follow \cite{KMO2},
where the light-cone dominance of the correlation functions containing one
heavy-quark field  was demonstrated.
Taking $q^2$ small but time-like, it is convenient
to consider the  rest frame of the virtual photon
with $q=(q_0,\vec{0}) $, defining
a unit vector $w=q/\sqrt{q^2}=(1,\vec{0})$, $w^2=1$.
The virtual $c$-quark fields  in (\ref{eq:Pi})
are then decomposed into the static part and residual 4-momentum:
$p_c=m_cw+\tilde{p}$, so that
$c(x)=\exp(-im_cw\cdot x)h_w(x) $, where the
effective field $h_w(x)$  contains only the $\tilde{p}$ components.
Note that the last redefinition only aims at separating the $m_c$ scale
from the virtual $c$-quark momentum and making the residual off-shell
momentum $\tilde{p}$ independent of this scale.
Since the virtual $\bar{c}$ quark 4-momentum is $p_{\bar{c}}=q-p_c=
q-(m_cw+\tilde{p})$, the $m_cw$ component of the Dirac-conjugated $\bar{c}$-field
enters with the same sign, so that $\bar{c}(x)=\exp(-im_cw\cdot x)\bar{h}_w(x)$.
After rescaling all $c$ and $\bar{c}$ fields,
the operator (\ref{eq:Pi}) transforms to
 \be
{\cal C}^a_{\mu}(q)=
\int d^4xe^{-i(2m_cw-q)\cdot x}
T \Big\{\bar{h}_w(x)\gamma_\mu h_w(x) ,
\bar{h}_{wL}(0)\Gamma^a h_{wL}(0) \Big\}\,,
\label{eq:Pi1}
 \ee
where the $m_c$-dependence
is now concentrated in the exponent. In the  above integral,
the dominant contribution stems from
the region where  the exponent
does not strongly oscillate, that is, where
$(2m_cw-q)\cdot x\sim 1$ ,
yielding
\be
\langle x^2\rangle\sim 1/(2m_cw-q)^2\sim 1/(2m_c-\sqrt{q^2})^2\,,
\label{eq:lcdom}
\ee
(for a more detailed derivation see e.g., \cite{CK}).
In the region $q^2\ll 4m_c^2$  the light-cone
OPE for the product of operators in (\ref{eq:Pi1}) can be applied.
When $q^2$ grows and approaches the threshold $4m^2_c$, the expansion
becomes invalid. Note that due to Lorentz-invariance,
the estimate (\ref{eq:lcdom}) is valid in any frame, including
the rest frame  of $B$-meson, where
both the $c$-quark pair and
$s$ quark  emitted in the heavy $b$-quark decay,  are
energetic.

\section{Expansion near the light-cone}

The expansion of  the operator-product
in (\ref{eq:Pi}) starts with contracting the $c$-quark fields
in the free propagators forming a two-point $c$-quark loop.
At leading order, there is no difference between the light-cone
and local OPE  and the hadronic matrix element is:
\ba
 \Big[{\cal H}^{(B\to  K^{(*)})}_\mu(p,q)\Big]_{fact}= \left(\frac{C_1}{3}+C_2\right)
 \langle  K^{(*)}(p)|{\cal O}_\mu(q)| B(p+q) \rangle \,,
\label{eq:Hmufact}
\ea
where both  $O_1$ and $O_2$ contribute and the local operator
\ba
 {\cal O}_\mu(q)=(q_\mu q_\rho- q^2g_{\mu\rho})
\frac{9}{32\pi^2}~g(m_c^2,q^2)\bar{s}_L\gamma^\rho b_L \,.
\label{eq:ope1} \ea is reduced to the $b\to s $  current. Hence
the matrix element (\ref{eq:Hmufact}) is factorized to  $B\to
K^{(*)}$  form factors. Diagrammatically, this contribution is
shown in Fig.~\ref{fig:diag}a. The charm-loop coefficient function
in (\ref{eq:ope1}) in the adopted operator basis is given by  the
well-known expression \cite{Heff}: \ba g(m_c^2,q^2)&=& -\frac89
\ln \left(\frac{m_c}{m_b}\right) + \frac{8}{27} +\frac{4}{9}y(q^2)
\nonumber\\
&& - \frac{4}{9}\left(2+y(q^2)\right)\sqrt{y(q^2)-1}
~\mbox{arctan}\left(\frac1{\sqrt{y(q^2)-1}}\right)\,,
\label{eq:loop} \ea where $y(q^2)= 4m_c^2/q^2 >1$, the
renormalization scale is taken as $m_b$ and $g(m_c^2,0)= -8/9
\ln(m_c/m_b)-4/9$. Furthermore, a dispersion relation for the
coefficient function in the variable $q^2$ is valid: \be
g(m_c^2,q^2) = g(m_c^2,0) + \frac{q^2}{\pi}\int \frac{\mbox{Im}_s
g(m_c^2,s)}{s(s- q^2)}ds \,. \label{eq:dloop} \ee with the
spectral density: \be \frac{1}{\pi}\,\mbox{Im}_s
g(m_c^2,s)=\frac4{9}\sqrt{1-\frac{4m_c^2}s}(1 + \frac{2m_c^2}s )
\Theta(s-4m_c^2)~. \label{eq:im1} \ee The factorizable amplitude
(\ref{eq:Hmufact}) is often called ``perturbative'' or
``short-distance''  charm-loop effect. There are gluon corrections
to this amplitude which do not violate factorization. For
instance, the perturbative gluon exchanges within the $c$-quark
loop  in Fig. {\ref{fig:diag}a are  known and can in principle be
added in a form of $O(\alpha_s)$ corrections to the coefficient
function $g(m_c^2,q^2)$. We will neglect them here. The
$O(\alpha_s)$ corrections to the weak-current vertex are
implicitly included in the calculation of the  hadronic form
factors, e.g., in LCSR  \cite{BZ04,DKMMO}.

The nonfactorizable contributions to the hadronic matrix element
 (\ref{eq:matr}) start with the one-gluon emission
from the charm-loop. Some of the nonfactorizable diagrams with
hard-gluon exchanges in $O(\alpha_s)$  are shown in
Fig.~\ref{fig:diag}c,d. They are included in the NLO factorization
formula \cite{BFS}, (see also \cite{BVNLO}) for $B\to
K^{(*)}\ell^+\ell^-$ or $B\to K^*\gamma$ in a form of the
hard-scattering kernels convoluted with the $B$- and $K^*$-meson
light-cone distribution amplitudes (DA's). These perturbative
effects can be taken into account separately and will not be
included in our analysis. Nonfactorizable two-gluon effects in
(\ref{eq:ope1}) with one hard and one soft gluon, or with two
hard gluons are also neglected, being presumably very small due
to extra $O(\alpha_s)$ suppression.

In this paper, we consider the emission
of one soft gluon (with low virtuality but nonvanishing momentum)
from the $c$-quark loop and evaluate the contribution of this mechanism to
the $B\to K^{(*)} \ell^+\ell^-$ amplitudes.
One of the corresponding diagrams is shown in Fig.~\ref{fig:diag}b,
the second one has a gluon emitted from the other $c$-quark line.

In terms of the light-cone expansion, the contraction of
$c$-quark fields entering (\ref{eq:matr}) yields a gluon-field operator
multiplied by a $b\to s$ colour-octet current.
Note that only the operator $O_1$ contributes at this level.
In order to calculate this contribution we use
the $c$-quark propagator near the light-cone \cite{BB}
including the one-gluon term:
\ba
\langle 0 \mid T\{c(x_1)\bar{c}(x_2)\}\mid 0 \rangle = -i\int \frac{d^4k}{(2\pi)^4}
e^{-ik(x_1-x_2)}\frac{\DS k+m_c}{m_c^2-k^2}\nonumber \\
-i\!\int\limits_0^1 du G^{\alpha\beta}(ux_1+\bar{u}x_2)\!\!\int \!\!\frac{d^4k}{(2\pi)^4}
e^{-ik(x_1-x_2)}
\frac{\bar{u}(\DS k+m_c)\sigma_{\alpha\beta}+u\sigma_{\alpha\beta}(\DS k+m_c)}
{2(m_c^2-k^2)^2}\,,
\label{eq:prop}
\ea
where $G^{\alpha\beta}(x)=g_s(\lambda^a/2)G^{a\,\alpha\beta}(x)$,
$\bar{u}=1-u$, and the fixed-point gauge
for the gluon field is used.
The above symmetric form can easily be derived from the
expression presented, e.g., in \cite{BBKR}.

To proceed, we define the light-cone kinematics in the rest-frame
of the decaying $B$-meson, introducing the unit vector
$v=p_B/m_B=(p_{K^{(*)}}+q)/m_B=(1,0,0,0)$, two light-cone vectors
$n_{\pm}$, so that :
\begin{eqnarray}
v &=& \frac{1}{2} (n_+ + n_-)\,, \qquad n_+^2 = n_-^2=0\,,
\qquad (n_+ n_-) = 2\,,\nonumber\\
q &=& (n_- q) {n_+ \over 2} +(n_+ q) {n_- \over 2}\,,
\end{eqnarray}
and choosing $\vec{q}_{\perp}=0$.
We consider a kinematical situation when the lepton-pair
invariant mass is small:
$q^2 = (n_+ q) (n_- q) \ll 4m_c^2<m_b^2$ , while
$(vq)=1/2[(n_+ q)+ (n_- q)]\sim m_b/2$ is large, so that
one of the components of $q$ in the above expansion dominates,
or equivalently, $q$ is approximately parallel to one of the light-cone
directions. We choose
\begin{equation}
(n_- q) \sim m_b, \quad   (n_+ q) \sim \frac{q^2}{ m_b} \ll (n_-q),
\quad \mbox{hence} \quad   q \simeq (n_- q) \frac{n_+}2\,.
\end{equation}

The propagator (\ref{eq:prop}) taken between $x$ and $0$
involves the gluon field at the point $ux$, $(0<u<1)$
which we rewrite via nonlocal differential operator:
\ba
G^{\alpha\beta} (ux) = \exp [ -i ux_\tau (i {\cal D}^\tau) ] \,\, G^{\alpha\beta}\,.
\label{eq:Gexp}
\ea
where $G^{\alpha\beta}=G^{\alpha\beta}(0)$. Note that due to the fixed-point gauge,
the simple derivative can be  replaced by the covariant derivative
acting on the gluon field and ensuring gauge invariance.

We are interested in the dominant effect of the nonvanishing gluon
momenta generated by the exponent in  (\ref{eq:Gexp}).
Decomposing the covariant derivative in the light-cone vectors
\be
{\cal D}= (n_+{\cal D})\frac{n_-}2+ (n_-{\cal D})\frac{n_+}2 +{\cal D}_{\perp},
\label{eq:Dexp}
\ee
we retain only the $n_-$ component, which corresponds
to the gluons emitted antiparallel to $q$, that
is, in the same direction as the $s$-quark in the $B$-meson  rest frame.
We then have
\begin{eqnarray}
G^{\alpha\beta} (ux) & \simeq &
 \exp [ -i  u(n_- x )  {(i n_+ {\cal D}) \over 2} ] G^{\alpha\beta}
\nonumber \\
&=& \int d\omega \, \exp [ -i u (n_- x)\omega ]  \, \delta [
\omega - {(i n_+ {\cal D}) \over 2}]  \,\,  G^{\alpha\beta} \,.
\end{eqnarray}

Inserting this expression in the propagator (\ref{eq:prop}),
and shifting the $k$ integration,
we obtain for the one-gluon part of the propagator:
\ba
 \langle 0 \mid T\{c(x)\bar{c}(0)\}\mid 0 \rangle_G
= -i \int d\omega \int\limits_0^1 du
\int \frac{d^4k}{(2\pi)^4}
\exp \left( -i k x \right)
\nonumber \\
\times \frac{\bar{u}(\DS k +m_c- u\DS n_-\omega)\sigma_{\alpha\beta} +
u \sigma_{\alpha\beta}(\DS k+m_c- u \DS n_-\omega)}{2(m_c^2-[k - u n_-\omega]^2)^2}
 \delta [ \omega -{ (i n_+ {\cal D})\over 2}]   G^{\alpha\beta} \,.
\label{eq:prop1}
\ea

Using this form of the propagator in (\ref{eq:matr}),
we cast the soft-gluon emission  part of the hadronic matrix element
in a form:
\ba
 \Big[{\cal H}^{(B\to  K^{(*)})}_\mu(p,q)\Big]_{nonfact}=
2 C_1\langle  K^{(*)}(p)|\widetilde{{\cal O}}_\mu(q)| B(p+q) \rangle \,,
\label{eq:Hmunonfact}
\ea
where $\widetilde{{\cal O}}_\mu(q)$ is
a convolution of the coefficient function with the nonlocal operator:
\begin{eqnarray}
\widetilde{{\cal O}}_\mu(q)
=  \int d\omega\, I_{\mu \rho \alpha\beta}(q,\omega)
\bar{s}_L\gamma^\rho
\delta [ \omega - {(i n_+ {\cal D}) \over 2}]
\widetilde{G}_{\alpha \beta} b_L \,\, .
\label{eq:nonloc}
\end{eqnarray}
In the above, $\widetilde{G}_{\alpha\beta}=\frac12\epsilon_{\alpha\beta\sigma\tau}
G^{\sigma\tau} $, and the derivative acts only on the gluon-field operator.
The coefficient function reads:
\ba
I_{\mu \rho \alpha\beta}(q,\omega)&=& {1 \over 8 \pi^2}
\int_0^1 du \Big \{
\Big[\bar{u}\tilde{q}_{\mu}\tilde{q}_{\alpha}g_{\rho\beta}
+u\tilde{q}_{\rho}\tilde{q}_{\alpha}g_{\mu\beta}
- \bar{u} \tilde{q}^2g_{\mu\alpha}g_{\rho\beta} \Big]
\frac{dI(\tilde{q}^2,m_c^2)}{d \tilde{q}^2}
\nonumber \\
&& -\frac{ \bar{u} - u}{2}
g_{\mu\alpha}g_{\rho\beta}I(\tilde{q}^2)
\Big \}\, ,
\label{eq:Wils}
\ea
where
\begin{eqnarray}
I(\tilde{q}^2,m_c^2)= \int_0^1 dt \ln \Big[\frac{\mu^2}{m_c^2 - t(1-t) \tilde{q}^2}\Big] \,,
\end{eqnarray}
is represented in a compact unintegrated form, and we use the notation
 $\tilde{q} = q -u\omega n_{-} $ , so that $\tilde{q}^2\simeq q^2-2u\omega m_b $.
Here we take into account that  $\omega\ll m_b$,
after the hadronic matrix element is taken.
Note that the neglected components of ${\cal D}$  in (\ref{eq:Dexp})
produce small, $O(\omega/m_b)$ corrections to $\tilde{q}^2$,
hence our approximation  is well justified.
The spectral density of this coefficient function has the following form:
\begin{eqnarray}
\frac{1}{\pi}{\rm Im} \, I_{\mu \rho \alpha\beta}(q,\omega)&=& {m_c^2 \over  4
\pi^2 \tilde{q}^2 \sqrt{\tilde{q}^2  (\tilde{q}^2 - 4 m_c^2)} }
\int_0^1 du \Big \{
\bar{u}\tilde{q}_{\mu}\tilde{q}_{\alpha}g_{\rho\beta}
+u\tilde{q}_{\rho}\tilde{q}_{\alpha}g_{\mu\beta} \nonumber \\
&& - \Big[ u
+ {(\bar{u} -u ) \tilde{q}^2 \over  4 m_c^2} \Big]
\tilde{q}^2g_{\mu\alpha}g_{\rho\beta} \Big
\}\Theta(\tilde{q}^2-4m_c^2) \,.
\label{eq:im2}
\end{eqnarray}

For a gluon field with fixed momentum one restores from (\ref{eq:nonloc}) and
(\ref{eq:Wils}) the expression
obtained in \cite{BIR} from the diagrams with the
on-shell gluon emission from the $c$-quark loop.
Since only the sum of the two diagrams
is ultraviolet-convergent, they have to be calculated in $D\neq 4$.
In the resulting expression, contrary to the comment
in \cite{BIR}, we found no mass-independent constant term.

Importantly, the operator (\ref{eq:nonloc}) has an overall
power-suppression factor $\sim 1/(4m_c^2-q^2)$ compensating the
gluon field-strength dimension. The $B\to K^{(*)}$ hadronic matrix
element of this operator is reduced to a specific ``nonlocal form
factor'' \be \langle K^{(*)}(p)|\bar{s}_L\gamma^\rho \delta [
\omega - {(i n_+ {\cal D}) \over 2}] \widetilde{G}_{\alpha \beta}
b_L| B(p+q)\rangle\,, \label{eq:Omatr} \ee which has to be
calculated with some nonperturbative method. In fact this matrix
element resembles a nonforward distribution with different initial
and final hadrons.

In the local OPE limit, that is, neglecting all derivatives of the gluon
field, (\ref{eq:nonloc}) is reduced to:
\ba
\widetilde{{\cal O}}^{(0)}_\mu(q)
= I_{\mu \rho \alpha\beta}^{(0)}(q)
\bar{s}_L\gamma^\rho \widetilde{G}_{\alpha \beta} b_L \,\, ,
\label{eq:Otilde0}
\ea
with
\ba
I^{(0)}_{\mu \rho \alpha \beta}(q)=I_{\mu \rho \alpha \beta}(q,0)=
(q_\mu q_\alpha g_{\rho\beta}+q_\rho q_\alpha
g_{\mu\beta}- q^2g_{\mu\alpha}g_{\rho\beta})
I^{(0)}(q^2,m_c^2)\,,
\label{eq:ope2}
\ea
and
\be
I^{(0)}(q^2,m_c^2)=
\frac1{16\pi^2}
\int_0^1 dt~ \frac{t(1-t)}{m_c^2-q^2t(1-t)} \,.\\
\label{eq:Pictil}
\ee
At $q^2=0$,  one easily recognizes  in (\ref{eq:Otilde0}) the
quark-gluon operator emerging from the
charm-loop with soft gluons \cite{Vol,KRSW}
(see also \cite{cloop,BIR}).
The 3-point  QCD sum rules in \cite{KRSW} and LCSR
in \cite{BZJ} were used to evaluate the
$B \to K^*\gamma$  hadronic matrix element of this operator.
The same approach was used in \cite{HF} to estimate
the nonfactorizable soft-gluon effect in $B\to J/\psi K$.

To clarify the difference between the light-cone and local OPE, we
return to the nonlocal operator (\ref{eq:nonloc}) and expand the
$\delta$-function in powers of the derivatives acting on the gluon
field. Integrating the coefficient function over $\omega$, one
encounters a tower of local operators: \be \widetilde{{\cal
O}}^{(n)}_\mu(q)= \frac{1}{n!}\frac{d^n}{d\omega^n}I_{\mu \rho
\sigma \beta}(q,\omega)\Big|_{\omega=0} \bar{s}_L\gamma^\rho
\big(\frac{in_+{\cal D}}{2}\big)^n\widetilde{G}_{\alpha \beta} b_L
\,\, . \label{eq:tower} \ee Sandwiching the $n$-th operator
between $B$ and $K^{(*)}$, one obtains a contribution of the order
of $(m_b\Lambda_{QCD})^n/(4m_c^2-q^2)^{n+1}$. Since $m_b
\Lambda_{QCD}$ is generally not smaller than $m_c^2$, there is no
parametric suppression of the $n\neq 0$ contributions, as already
discussed in \cite{cloop}. More recently, the description of the
charm-loop contribution in  $B\to X_s\gamma $ in terms of nonlocal
operators was advocated in \cite{LNP}. In what follows, we apply
LCSR to evaluate the hadronic matrix elements (\ref{eq:Omatr}),
thereby performing an effective resummation of the local operators
in (\ref{eq:tower}).

Expanding the $c$-quark propagator near the light-cone \cite{BB} further,
one encounters higher-dimensional nonlocal operators with two and
more gluon fields,  each of them
generating a tower of operators with calculable, albeit
complicated coefficient functions.
Simple dimensional argument based on the presence of extra gluon fields
in $B$-meson DA's allows one to anticipate an overall
suppression of these two and more soft-gluon contributions
in LCSR by additional powers of the scale $1/(4m_c^2-q^2)$ with  respect  to
the leading one-gluon term.

In exclusive $B$ decays, the soft-gluon effects stemming
from the light-cone expansion in the framework of LCSR were considered
in the analysis of $B\to \pi\pi$ in \cite{AKBpipi}
and  $B\to J/\psi K$ in \cite{BM}, where
the LCSR with DA's of light mesons were used. In this case
as shown in \cite{AKBpipi}, an artificial four-momentum in the
vertex of the weak operator has to be introduced. In the following section
we use a simpler approach where this modification can be avoided.

\section{LCSR for hadronic matrix elements}

The $B\to K^{(*)}$ transition matrix elements (\ref{eq:matr})
determining the charm-loop effect   can now be represented
as a sum of  the hadronic matrix elements of the effective
operators ${\cal O}_\mu$ and $\widetilde{{\cal O}}_\mu$, containing
the factorizable and (soft) nonfactorizable contributions, respectively.

Let us first consider the $B\to K$ transition. Adding
(\ref{eq:Hmufact}) and (\ref{eq:Hmunonfact}) together, we obtain:
\ba {\cal H}_\mu^{(B\to K)}(p,q)
&=&\left(\frac{C_1}3+C_2\right)\langle K(p)| {\cal O}_\mu(q)
|B(p+q)\rangle
+2C_1\langle K(p)| \widetilde{{\cal O}}_\mu(q) |B(p+q)\rangle
\nonumber \\
&=&\big[(p\cdot q)q_\mu-q^2p_\mu\big]{\cal H}^{(B\to K)}(q^2)\,,
\label{eq:Hmu} \ea where  the e.m. current conservation is taken
into account and \be {\cal H}^{(B\to K)}(q^2)=
\left(\frac{C_1}3+C_2\right)\A(q^2)+2C_1\tilde{\A}(q^2)
\label{eq:HBK} \ee contains two invariant amplitudes
parameterizing the two hadronic matrix elements in (\ref{eq:Hmu}).
\FIGURE[t]{
\includegraphics[scale=0.80]{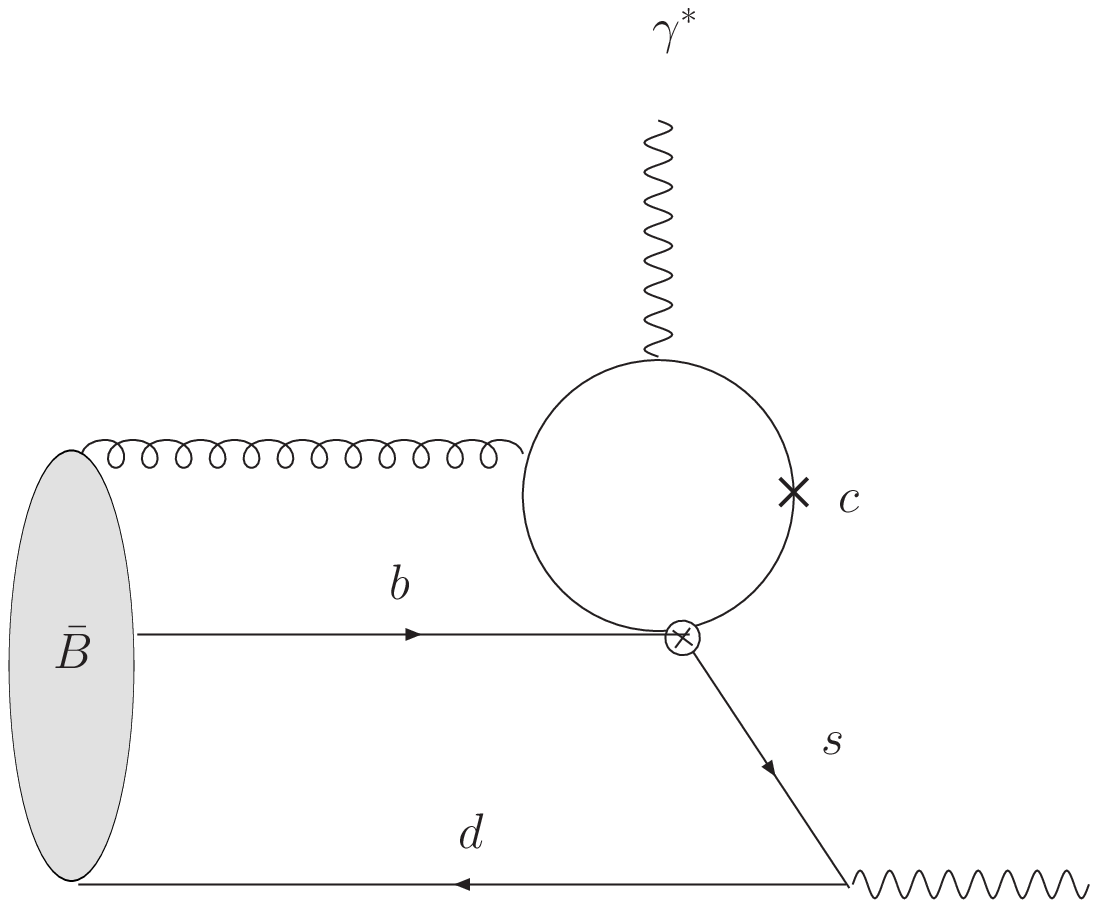}
\caption{\it The correlation function used to calculate the $B\to
K^{(*)}$ matrix element of the soft-gluon emission from the charm
loop. The oval denotes the $B$-meson quark-antiquark-gluon
distribution amplitudes. The $c$-quark loop with the emitted gluon
generates the nonlocal effective operator $\widetilde{O}_\mu$. The
cross indicates the point of gluon emission in  the second
diagram. }
\label{fig:cloop}
}

In the adopted approximation, the amplitude $\A$ is factorized:
\begin{eqnarray}
\A(q^2) =
\frac{9}{32\pi^2}~g(m_c^2,q^2)f^{+}_{BK} (q^2) \,,
\label{eq:Afact}
\end{eqnarray}
if one uses (\ref{eq:ope1})
and the standard definition of the $B\to K$ form factor:
\be
\langle K (p)| \bar{s}_{L}\gamma_{\rho}b_L  |
B(p+q) \rangle =
f^+_{BK}(q^2)p_{\rho}+...
\label{eq:BKformf}
\ee

Our main task is to calculate the amplitude
$\tilde \A $ determining the soft-gluon emission from the
charm-loop. To this end, we employ LCSR
with the $B$-meson DA's and, following \cite{KMO1,KMO2},
introduce the correlation function:
\begin{eqnarray}
{\cal F}^{(B\to K)}_{\nu \mu}(p,q)=i\int d^4y e^{ip\cdot y} \langle 0| T\{
j_{\nu}^{K}(y) \widetilde{{\cal O}}_\mu(q)\}| B(p+q)\rangle \,,
\label{eq:BKcorr}
\end{eqnarray}
where $j_{\nu}^{K} = \bar{d}\gamma_{\nu} \gamma_5 s$ is
the $K$-meson interpolating current and
$B$ meson is taken on shell, as a HQET state: $ | B(p+q)\rangle\simeq
|B(v)\rangle $.
Inserting the full set of states with the kaon
quantum numbers
between the current $j_{\nu}^{K}$ and $\widetilde{{\cal O}}_\mu$ in
(\ref{eq:BKcorr}) we obtain the hadronic dispersion relation:
\begin{eqnarray}
{\cal F}^{(B\to K)}_{\nu \mu}(p,q) =  {i f_{K} p_{\nu} \over
m_K^2-p^2 } [ (p \cdot q) q_{\mu}-q^2p_{\mu} ]  \tilde{\A}(q^2) +
\int_{s_h}^{\infty} ds\,\, { \tilde{\rho}_{\nu \mu }(s, q^2) \over
s-p^2}\, , \label{eq:BKcorr1}
\end{eqnarray}
where $f_K$ is the kaon decay constant defined as $\langle 0 |
\bar{d}\gamma_{\nu} \gamma_5 s | K(p)\rangle = i f_K p_{\nu}$ and
the spectral density $\tilde{\rho}_{ \nu \mu}(s, q^2)$
accumulates excited and continuum states
with the kaon quantum numbers, located above the threshold
$s_h$.

Two comments are in order. First, in the approach we are using,
hadronic matrix elements are related to the correlation function
via dispersion relation. Hence, the ``full" hadronic matrix
element of $B\to K$ transition with the soft-gluon emission enters
the residue of the kaon pole in (\ref{eq:BKcorr1}). In other
words, although we have chosen a particular correlation function
with $B$-meson DA's, there is no need to add a contribution where
a soft gluon emitted from the charm loop enters the final-state
kaon ``wave function". Secondly,  at very large timelike $p^2$
there are also ``parasitic" charm-anticharm states contributing to
the hadronic spectral density in (\ref{eq:BKcorr1}), but they are
heavily suppressed after the Borel transformation in $p^2$. This
circumstance allows one to avoid introducing an auxiliary
4-momentum in the effective-operator vertex, as suggested in
\cite{AKBpipi}.

In \cite{KMO2} the form factor $f^{+}_{BK} (q^2)$ was calculated
from the correlation function  similar  to (\ref{eq:BKcorr}),
where, instead of the complicated effective operator, the $b\to
s$, vector current was inserted. LCSR was obtained including the
contributions of two-particle (quark-antiquark) and three-particle
(quark-antiquark-gluon) $B$-meson DA's. Here the leading-order
diagrams shown in Fig.~\ref{fig:cloop} involve only the
three-particle DA's of $B$ meson. We calculate these diagrams
contracting the $s$-quark fields. The result reduces to the
vacuum-to-$B$ matrix element of the $\bar{d}G b$ nonlocal
operator. It is decomposed in HQET in four three-particle
$B$-meson DA's \cite{Kawamura}: \ba && \hspace{-0.5 cm}  \langle
0| \bar{d}_{\alpha}(y) \delta [ \omega - {(i n_+ {\cal D}) \over
2}] G_{\sigma \tau}(0) b_{\beta} (0)| \bar{B}(v)\rangle
 \\
&& \hspace{-0.5 cm} = { f_B m_B \over 2} \int_0^{\infty} d \lambda
\, e^{-i\lambda y\cdot v} \bigg [  (1+ \not v) \bigg \{
(v_{\sigma} \gamma_{\tau} - v_{\tau} \gamma_{\sigma} )
 \big[\Psi_A(\lambda,2\omega)-\Psi_V(\lambda,2\omega)\big]
\nonumber \\&& \hspace{-0.5 cm} - i \sigma_{\sigma
\tau}\Psi_V(\lambda,2\omega)
 -{ y_{\sigma} v_{\tau} - y_{\tau} v_{\sigma} \over v \cdot y}
X_A(\lambda, 2\omega) + { y_{\sigma} \gamma_{\tau} -
y_{\tau}\gamma_{\sigma} \over v \cdot y} Y_A(\lambda,2\omega )
\bigg \}\gamma_5\bigg]_{\beta\alpha} \,, \nonumber \label{eq:BDAs}
\ea where $f_B$ and $m_B$ are the $B$-meson decay constant  and
mass, respectively. Further details can be found in \cite{KMO2}.

Equating the correlation function $\cal{F}_{\nu\mu}$, written in
terms of the $B$-meson DA's, to  its hadronic representation
(\ref{eq:BKcorr1}), we perform the Borel transformation in the
variable $p^2$ and employ quark-hadron duality in the kaon channel
to approximate the integral over the spectral density
$\tilde{\rho}_{\nu\mu}$, introducing the effective duality
threshold $s_0^K$. The resulting LCSR for the nonfactorizable
$B\to K$ hadronic matrix element  of the charm-loop effect reads:

\ba \mbox{}\hspace{-1cm} && \tilde{\A}(q^2)= -\frac{f_B
m_B}{8\pi^2 f_K(m_B^2-m_K^2-q^2)} \int\limits_0^{\sigma_0} d\sigma
\int\limits_0^\infty \! \!d\omega \int\limits_0^1
\!\!du\int\limits_0^1\!\!dt \, \exp\Big(\frac{m_K^2-\sigma m_B^2}{M^2}\Big)  \nonumber\\
&&  \times \!\frac{1}{m_c^2-t(1-t)(q^2-2m_B u\,\omega)}
\Bigg[C^{(\Psi_V)}(q^2,u,\sigma,\omega,t)\Psi_V(m_B\sigma,2\omega)
\nonumber\\
&& +C^{(\Psi_{AV})}(q^2,u,\sigma,\omega,t)\big[\Psi_V(m_B\sigma,
2\omega)- \Psi_A(m_B\sigma, 2\omega)\big]
\\
&& +C^{(X_A)}(q^2,u,\sigma,\omega,t)\bar{X}_A(m_B\sigma, 2\omega)
-\frac{d}{d\sigma}\big[
\widetilde{C}^{(X_A)}(q^2,u,\sigma,\omega,t)\bar{X}_A(m_B\sigma,
2\omega) \big]
\nonumber\\
&& +C^{(Y_A)}(q^2,u,\sigma,\omega,t)\bar{Y}_A(m_B\sigma, 2\omega)
-\frac{d}{d\sigma}\big[\widetilde{C}^{(Y_A)}(q^2,u,\sigma,\omega,t)\bar{Y}_A(m_B\sigma,
2\omega) \big] \Bigg], \nonumber
 \label{eq:LCSRBK}
\ea
where $M^2$ is the Borel parameter, $\sigma_0\simeq
s^K_0/m_B^2$ (up to small corrections), \ba \bar{X}_A
(\lambda,\omega) = \int_0^{\lambda} d \tau  X_A(\tau,\omega),
\qquad  \bar{Y}_A (\lambda, \omega) = \int_0^{\lambda} d \tau
Y_A(\tau,\omega) . \ea and the coefficients $C^{(...)}$ and
$\widetilde{C}^{(...)}$ are collected in Appendix~D. The sum rule
is presented in (\ref{eq:LCSRBK})  in a simplified form, omitting
the $s$-quark mass and small $O(1/m_B^2)$ corrections. In the
numerical analysis the complete expression is used.

Note that the spectral density of the correlation function
(\ref{eq:BKcorr}) in the variable $p^2$, in addition to the $s$-quark pole,
contains also ``parasitic'' contributions which correspond to
putting simultaneously on-shell the $s$-quark and the $\bar{c}c$-quark lines
 in the diagram of Fig.~\ref{fig:cloop}.
After Borel transformation these terms are suppressed by $\sim
\exp(-4m_c^2/M^2)$ and we neglect them. As noted above, the
corresponding hadronic states are also neglected in the
dispersion relation.

In order to assess the accuracy of using the nonlocal
effective operator $\tilde{\cal O}_\mu$ in the correlation function
we derived the LCSR in an alternative way,
inserting the operator $C_1O_1+C_2O_2$ and the c-quark e.m. current
directly in the correlation function. As a result we obtain a
more complicated expression for LCSR,
which differs from  (\ref{eq:LCSRBK}) only by terms
suppressed  by inverse powers of $m_c$ and/or $m_B$,
and yields numerically very close results for the amplitude $\tilde{\A}$.
Finally, the local OPE limit of LCSR  (\ref{eq:LCSRBK}) was also investigated,
that is, when the soft-gluon momentum is neglected. This limit literally
corresponds to putting $\omega\to 0$ in the coefficients $C^{(...)}$
and in the denominator of (\ref{eq:LCSRBK}). The numerical
influence of this approximation  will  be discussed in the next section.

Turning to the calculation of the $B\to K^*$ transition matrix
element defined in (\ref{eq:matr}) we decompose it into the three
kinematical structures: \ba {\cal H} _\mu^{(B\to K^*)}(p,q) &=&
\left(C_2+\frac{C_1}3\right) \langle K^*(p)| {\cal O}_\mu(q)
|B(p+q)\rangle
+2C_1 \langle K^*(p)| \widetilde{{\cal O}}_\mu(q)
|B(p+q)\rangle
\nonumber\\
&=& \epsilon_{\mu \alpha \beta \gamma } \epsilon^{\ast \alpha }
q^{\beta}  p^{\gamma} {\cal H}_1(q^2) +i[(m_B^2-
m^2_{K^{\ast}})\epsilon^{\ast}_\mu
-(\epsilon^\ast \!\cdot q) (2p+q)_{\mu}] {\cal H}_2(q^2) \nonumber \\
&& + i (\epsilon^\ast\! \cdot q) \bigg[q_{\mu} - {q^2 \over m_B^2-
m^2_{K^{\ast}}} (2 p+q)_{\mu} \bigg] {\cal H}_3(q^2)\,,
\label{eq:Kstampl} \ea where $\epsilon$ is the polarization vector
of the $K^*$-meson.

The invariant amplitudes in the above decomposition contain
factorizable and nonfactorizable parts stemming from the
matrix elements of ${\cal O}_\mu$ and $\widetilde{{\cal O}}_\mu$,
respectively:
\ba
{\cal H}_i^{(B\to K^*)}(q^2)= \left(C_2+\frac{C_1}3\right)\V_i(q^2)
+2C_1\widetilde{\V}_i(q^2)\,~~~ (i=1,2,3).
\label{eq:HBKst}
\ea

The three factorizable amplitudes $\V_i$
are easily obtained if one uses (\ref{eq:ope1})
and the standard definition of $B\to K^*$  vector and axial-vector form factors:
\ba
2\langle K^{\ast}(p)|\bar{s}_L\gamma_\rho b_L|B(p+q)\rangle
=\epsilon_{\rho\alpha\beta\gamma}
\epsilon^{*\alpha}q^\beta p^\gamma\dfrac{2 V^{BK^*}(q^2)}{m_B+m_{K^*}}
\nonumber\\
-i\epsilon_\rho^*(m_B+m_{K^*})A_1^{BK^*}(q^2)
+i(2p+q)_\rho(\epsilon^*q)\dfrac{A_2^{BK^*}(q^2)}{m_B+m_{K^*}}
+\ldots \,,
\label{eq:BKstff}
\ea
where the form factors  multiplying $q_\rho$ do not contribute
and are indicated by ellipses.
We have:
\ba
  \V_1(q^2) &=& -{9q^2
\over 32\pi^2(m_B + m_{K^{\ast}})} g(m_c^2,q^2) V^{BK^{\ast}} (q^2) \,,
\nonumber \\
 \V_2(q^2) &=&  {9q^2
\over 64\pi^2(m_B - m_{K^{\ast}})}  g(m_c^2,q^2) A_1^{BK^{\ast}}
(q^2)\, ,
\nonumber \\
\V_3(q^2) &=&  \frac{9(m_B^2 -
m_{K^{\ast}}^2)}{64\pi^2}g(m_c^2,q^2)\Bigg(\frac{A_2^{BK^{\ast}}(q^2)}{m_B+
m_{K^{\ast}}}-\frac{A_1^{BK^{\ast}}(q^2)}{m_B-m_{K^{\ast}}}\Bigg)\,.
\label{eq:Vfact} \ea

To calculate the  nonfactorizable amplitudes
$\widetilde{\V}_i(q^2)$ ($i=1,2,3$), describing the soft-gluon
emission we again resort to the LCSR method, introducing the
correlation function \ba {\cal F}^{(B\to K^*)}_{\nu \mu}(p,q) =i
\int d^4 y \,e^{i p \cdot y} \langle 0 | T \{
j_{\nu}^{K^{\ast}}(y) \widetilde{{\cal O}}_\mu(q) \}|B(p+q)\rangle
\,, \label{eq:corrKstar} \ea where $j_{\nu}^{K^{\ast}} = \bar{d}
\gamma_{\nu}s$ is the interpolating current for the $K^{\ast}$
meson. The hadronic dispersion relation for this correlation
function reads: \ba {\cal F}^{(B\to K^*)}_{\nu \mu}(p,q) =
\frac{f_{K^*}m_{K^*} \overline{\epsilon_{\nu}\langle K^*(p)|}
\widetilde{{\cal O}}_\mu(q) |B(p+q)\rangle }{ m^2_{K^{*}}-p^2 } +
\int_{s_h^*}^{\infty} ds\,\, { \widetilde{\rho}\,^*_{\nu \mu}(s,
q^2) \over s-p^2}\, , \label{eq:BKsthadr} \ea where $f_{K^*}$ is
the $K^*$  decay constant defined as $\langle 0 |
j_{\nu}^{K^{\ast}} | K^*(p)\rangle = \epsilon_\nu m_{K^*}f_{K^*}$
and the overline denotes the average over $K^*$ polarizations. We
neglect the total width of $K^*$. This approximation can in
principle be avoided by introducing a Breit-Wigner type
parameterization for this resonance. Excited and continuum states
with $K^*$ quantum numbers above the threshold $s_h^*$ contribute
to the spectral density $\tilde{\rho}^*_{\nu \mu}(s, q^2)$. The
latter is approximated using quark-hadron duality and introducing
an effective threshold $s_0^{K^*}$.

Furthermore, the $K^*$ contribution in (\ref{eq:BKsthadr}) is
written in terms of invariant amplitudes: \ba \big[{\cal F}^{(B\to
K^*)}_{\nu \mu}(p,q)\big]^{K^*}\!\!\!\! &=&\!\!\!\! \,\,
\frac{if_{K^*}m_{K^*}}{m_{K^{*2}}-p^2}\Bigg[ i \epsilon_{\mu \nu
\beta \gamma }q^{\beta}p^{\gamma} \tilde\V_1(q^2)
- (m_B^2-m^2_{K^{\ast}})g_{\mu\nu}\tilde\V_2(q^2)
\nonumber\\
\!\!\!\!&+&\!\!\!\! \,\, q_\mu q_\nu
\bigg(\tilde\V_2(q^2)-(1-\frac{q^2}{m_B^2-m_{K^{*2}}})
\tilde\V_3(q^2)\bigg)\Bigg] + ...\,, \label{eq:decompKst} \ea
 where we only show the kinematical
structures that have been used in our analysis. Applying the same
decomposition to the OPE result for the correlation function
${\cal F}^*_{\nu \mu}$ we obtain three sum-rule relations which
are then used to calculate the hadronic matrix elements
$\tilde{\V}_{1,2,3}(q^2)$. The derivation of these LCSR in terms
of $B$-meson DA's is very similar to the one described for $B\to
K$ case. Indeed, the only difference between the two correlation
functions (\ref{eq:BKcorr}) and (\ref{eq:corrKstar}) is in the
quantum numbers of the $s$-quark current. The resulting LCSR for
$\tilde{\V}_i(q^2)$ has the same structure as (\ref{eq:LCSRBK}),
only the coefficients multiplying the $B$-meson DA's and the
overall normalization factors are different. In order not to
overload this paper, we do not present these expressions here.

\section{Numerical analysis }

We use LCSR
with $B$ meson DA's to
evaluate not only the nonfactorizable amplitudes  $\widetilde{\A}$ and $\widetilde{\V_i}$
but also the $B\to K$ and $B\to K^*$ form factors
entering the factorizable parts $\A$ and $\V_i$ given
by (\ref{eq:Afact}) and (\ref{eq:Vfact}).
The sum rules for these form factors and their input
are taken from \cite{KMO2},
hence we use the same input for the new LCSR obtained in the
previous section.
In particular, for the three-particle $B$-meson
DA's the model suggested in \cite{KMO2} is taken:
\ba
\Psi_A(\lambda,\omega)&=&\Psi_V(\lambda,\omega)= { \lambda_E^2 \over 6
\omega_0^4} \omega^2 e^{-(\lambda+\omega)/ \omega_0} \, , \nonumber \\
X_A(\lambda,\omega) &=& { \lambda_E^2 \over 6 \omega_0^4} \omega (2
\lambda -\omega) e^{-(\lambda+\omega) /\omega_0} \, , \nonumber \\
Y_A(\lambda,\omega) &=& - { \lambda_E^2 \over 24 \omega_0^4} \omega (7
\omega_0 -13 \lambda + 3\omega) e^{-(\lambda+\omega) /\omega_0} \,.
\label{eq:3partDA}
\ea
In this model the parameter $\omega_0$ is equal to the
inverse moment $\lambda_B$ of the $B$ meson two-particle DA
$\phi^B_{+}$
and the normalization constant of the three-particle
DA's is $\lambda^2_E=3/2\lambda_B^2$.
For the inverse moment we use the interval obtained in
\cite{BIK} from QCD sum rule in HQET:
$\lambda_B (1 {\rm GeV}) = 460 \pm 110 \, {\rm MeV}\,$.
The scale-dependence of this parameter is neglected. To
be consistent with the $O(\alpha_s)$ accuracy of $\lambda_B$, the
$B$-meson decay constant $f_B =180 \pm 30 ~~{\rm MeV}$ obtained from
the two-point  QCD sum rules in $O(\alpha_s)$  is used.
Furthermore, the decay
constants and threshold parameters of $K$ and  $K^{\ast}$ mesons
are taken the same as in \cite{KMO2}:
\ba
f_{K} &=& 159.8  \pm 1.4 \pm 0.44 \, {\rm MeV} \, , \qquad
f_{K^{\ast}} = 217  \pm 5 \, {\rm MeV} \,, \nonumber \\
s_0^{K} &=& 1.05 \, {\rm GeV^2} \, , \qquad \hspace{2 cm}
s_0^{K^{\ast}} = 1.7 \, {\rm GeV^2} \, .
\label{eq:fconst}
\ea
For the Borel parameter interval in LCSR
we use is $M^2= 1.0\pm 0.25~{\rm GeV^2}$, slightly narrower
than the interval in \cite{KMO2}.

In addition, we need the $c$-quark mass value.
Since we are dealing with virtual $\bar{c}c$-quark loops, it is natural
to employ the $\overline{MS}$ mass,
assuming  that the normalization
scale is about $2\bar{m}_c$. As a default value,
we take $\bar{m}_c(2 \bar{m}_c)=1.05 ~{\rm GeV}$,
rescaled from the central value of  $\bar{m}_c$
obtained from charmonium sum rule in \cite{Kuhn}. To assess the
related uncertainty
we allow the normalization scale to change between $2m_c$ and $m_c$,
shifting the $c$-quark mass to  $m_c(\bar{m}_c)=1.30~\mbox{GeV}$.
Finally, $m_s(\rm 2 \,GeV) = 98 \pm 16~{\rm MeV}$ is adopted
(the average of nonlattice determinations \cite{ms}).

\FIGURE[t]{
\includegraphics[scale=0.8]{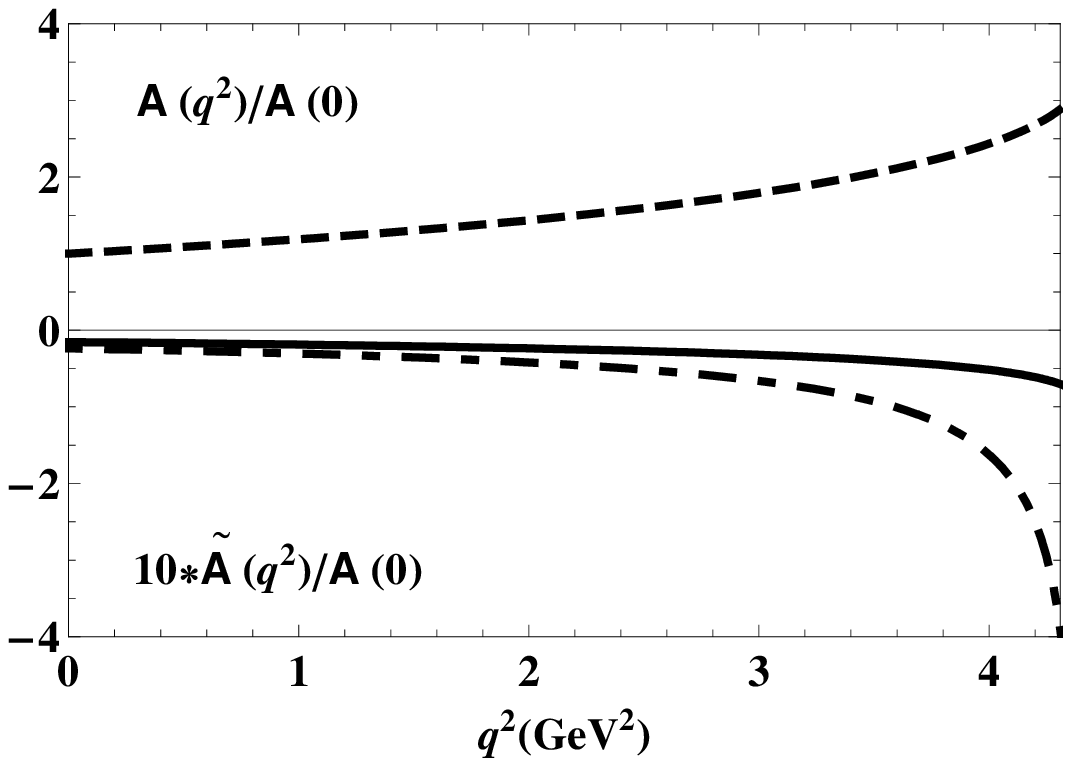}
\caption{\it The nonfactorizable $B\to K$ amplitude
$\widetilde{\A}(q^2)$ of the nonlocal effective operator
$\widetilde{O}_\mu$ (solid) and its local limit (dash-dotted),
plotted as a function of $q^2$, together with the factorizable
amplitude $\A(q^2)$ (dashed). The soft-gluon contributions are
rescaled with a factor of 10.  }
\label{fig:locnonloc}
}

In Fig.~\ref{fig:locnonloc} our numerical results for the
$q^2$-dependence of the dimensionless hadronic amplitudes
$\A(q^2)$ and $\widetilde{\A}(q^2)$
are displayed  for the central values of the input.
For convenience, we normalize both amplitudes
to $\A(0)=7.3\times 10^{-3}$.
The nonfactorizable amplitude is about a few percent of the
factorizable one and has the opposite sign.
Note that both $\A(q^2)$ and $\widetilde{\A}(q^2)$
develop imaginary parts above $4m_c^2$.
In the same figure we plot the nonfactorizable amplitude
$\widetilde{\A}(q^2)$ obtained in the  local OPE limit.
In this case, the nonfactorizable effect changes
substantially and diverges approaching to $q^2=4m_c^2$.
We come to an important conclusion that the effective resummation
the local operators in the framework of the light-cone OPE
``softens'' the  gluon correction to the charm loop.
\FIGURE[h]{
\includegraphics[scale=0.9]{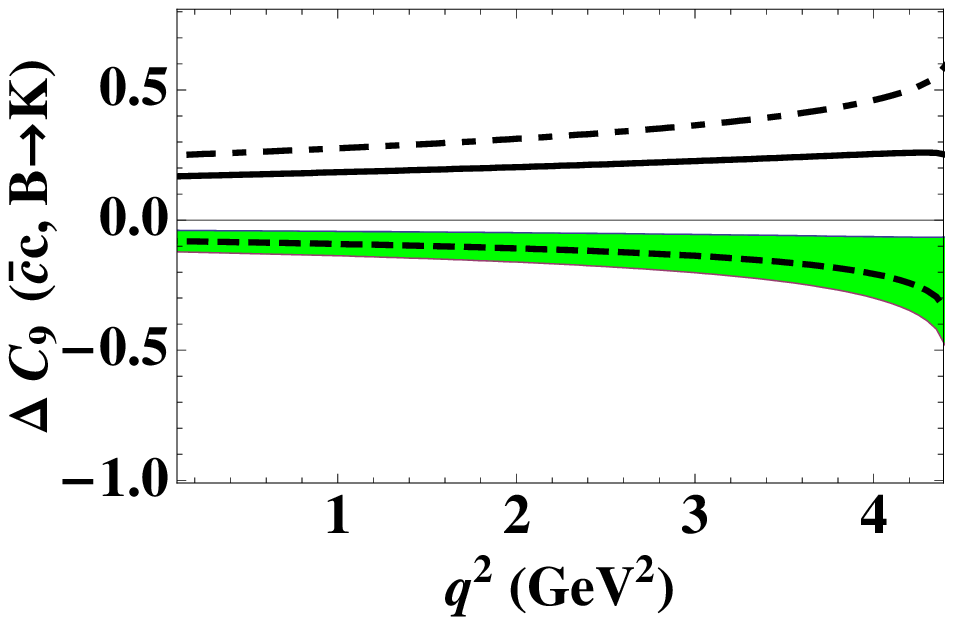}
\caption{\it The charm-loop  effect in $B\to K \ell^+\ell^-$
expressed as a correction to the Wilson coefficient $C_9$ (solid),
including  the nonfactorizable soft-gluon contribution (dashed)
with the shaded region indicating the estimated uncertainty and
the factorizable contribution (dash-dotted). }
\label{fig:BKres}
}

In the decay amplitude the soft-gluon  contribution
to the charm-loop effect gets enhanced considerably with respect to the
factorizable one by the ratio of the Wilson coefficients
$ 2C_1/(C_2+C_1/3)\gg 1$. Not surprisingly then, the proportion of the two
contributions depends on the normalization scale of
the Wilson coefficients. The scale can be fixed more accurately
when the perturbative  gluon corrections in the hadronic matrix
elements are taken into account, which is beyond our approximation.
\FIGURE[h]{
\hspace{1.5 cm}\includegraphics[scale=0.7]{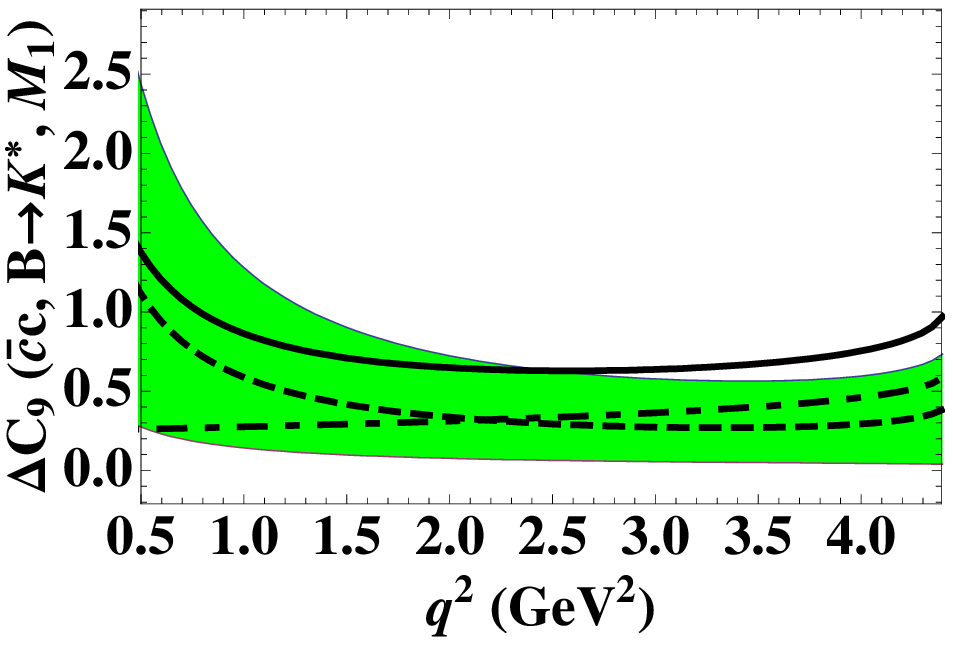} \\
 \includegraphics[scale=0.75]{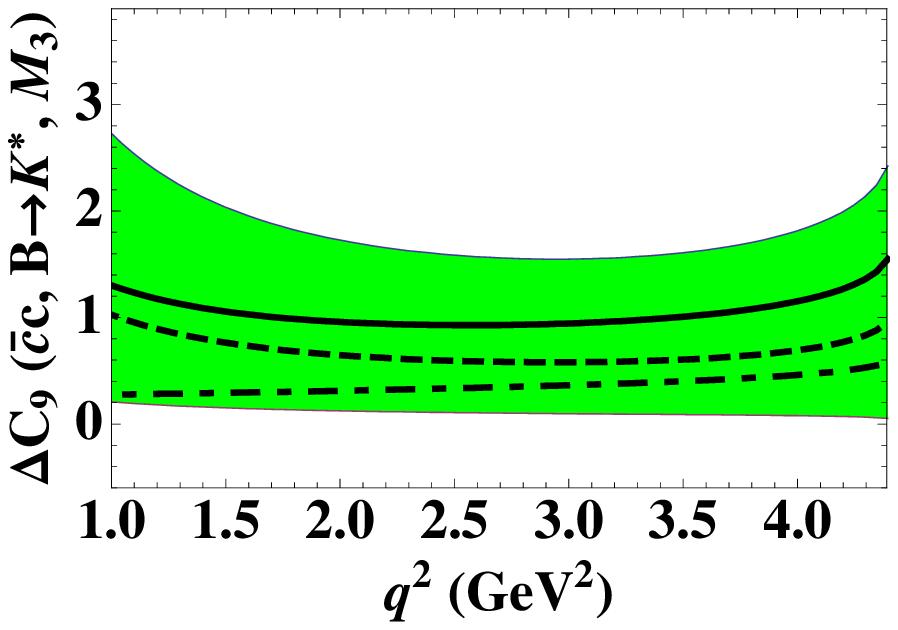}
\caption{The same as in Fig.~\ref{fig:BKres}
 for $\bar{B}_0 \to \bar{K^{\ast}} l^{+} l^{-}$.
The results for the correction to the amplitude ${\cal M}_2$ are
almost indistinguishable from the ones for ${\cal M}_1$ shown in
the upper panel.} \label{fig:BKstres}
}
For the numerical estimates we use the values of the Wilson coefficients
given  in Appendix~A. They are calculated in leading log approximation at
$\mu=m_b$, allowing the scale to vary
between $0.5m_b$ and $1.5m_b$. Note that in our calculation
the $b$-quark mass value only enters the effective Hamiltonian.
We adopt the $\overline{MS}$ value  extracted from
the bottomonium sum rules in \cite{Kuhn}
$\bar{m}_b(\bar{m}_b) = 4.164 \pm 0.025 ~{\rm GeV}$
(conservatively doubling the error).

It is convenient to express the charm-loop contribution to the
$B\to K \ell^+\ell^-$ amplitude in a form of a (process- and
$q^2$-dependent) correction to the Wilson coefficient $C_9$: \ba
C_9\to C_9+\Delta C_9^{(\bar{c}c,B\to K)}(q^2)\,. \ea Substituting
(\ref{eq:Hmu}),(\ref{eq:HBK}) and (\ref{eq:Afact}) to
(\ref{eq:ampl}) and comparing the result with the contribution of
$O_9$ to the $B\to K \ell^+\ell^-$ amplitude given in Appendix A,
we obtain: \ba \Delta C_9^{(\bar{c}c,B\to K)}(q^2)
&=&\frac{32\pi^2}{3}\frac{{\cal H}^{(B\to K)}
(q^2)}{f^+_{BK}(q^2)}
\nonumber\\
&=&\left(C_1+3C_2 \right)g(m_c^2,q^2)
+2C_1\widetilde{g}^{(\bar{c}c,B\to K)}(q^2)\,, \label{eq:deltC9}
\ea where the function \be \widetilde{g}^{(\bar{c}c,B\to
K)}(q^2)=\frac{32\pi^2}{3} \frac{\tilde{\A}(q^2)}{f^{+}_{BK}(q^2)}
\label{eq:deltCBK} \ee determines the new soft-gluon
nonfactorizable part of $\Delta C_9^{(\bar{c}c,B\to K)}$  and
represents our main result. Here only the ratio of the calculated
hadronic matrix elements enter and they both are calculated within
one and the same LCSR approach. The correction $\Delta
C_9^{(\bar{c}c,B\to K)}$ and its factorizable and nonfactorizable
parts are plotted in Fig.~\ref{fig:BKres} where the uncertainties
stemming from our calculation of $\widetilde{g}^{(\bar{c}c,B\to
K)}(q^2)$ are indicated. We vary all input parameters within their
adopted intervals and add  individual variations in the
quadrature. In Table \ref{tab:gtilde} we display the value and the
estimated uncertainties of  $\widetilde{g}^{(\bar{c}c,B\to K)}$ at
$q^2=1$ GeV$^2$, except the variation due to uncertainty of $m_s$
which is negligibly  small.

\TABLE[b]{
\caption{\it The functions determining the soft-gluon
correction to $C_9$, calculated from LCSR, central   values at
$q^2=1 \, \rm GeV^2$, and the uncertainties $\Delta_a$ caused by
the variations of the input parameters ($\delta_{m_c}= +0.25 \,
{\rm GeV}$, $\delta_{M^2}=^{+0.25}_{-0.25} \, {\rm GeV}$,
$\delta_{\lambda_B}= ^{+110}_{-110} \, {\rm MeV}$). }
\begin{tabular}{|c|c|c|c|c|}
  \hline&&&&\\[-3mm]
function & $\tilde{g}^{(\bar{c} c, B \to K)}$  &
   $\tilde{g}^{(\bar{c} c, B \to K^{\ast}, M_1)}$ &
   $\tilde{g}^{(\bar{c} c, B \to K^{\ast}, M_2)}$ &
    $\tilde{g}^{(\bar{c} c, B \to K^{\ast}, M_3)}$ \\
  \hline&&&&\\[-3mm]
centr.value & $-0.041$ & $0.26$ & $0.27$ & $0.46$ \\
 \hline&&&&\\[-3mm]
 $\Delta_{m_c}$ & $+0.014$ & -0.08 & -0.09 & -0.15  \\
\hline&&&&\\[-3mm]
$\Delta_{M^2}$ &  $^{+0.00}_{-0.001}$ &
$^{-0.04}_{+0.07}$ &  $^{-0.04}_{+0.08}$ &  $^{-0.07}_{+0.12}$ \\[1.5mm]
\hline&&&&\\[-3mm]
$\Delta_{\lambda_B }$ & $^{-0.016}_{+0.017}$ &
$^{+0.30}_{-0.17}$ & $^{+0.36}_{-0.18}$ & $^{+0.75}_{-0.33}$\\[1mm]
\hline&&&&\\[-3mm]
$\Delta_{tot}$  & $^{+0.022}_{-0.016}$ &  $^{+0.31}_{-0.19}$ &
$^{+0.37}_{-0.21}$  & $^{+0.76}_{-0.37}$ \\[1mm]
  \hline
\end{tabular}
\label{tab:gtilde}
}

Note that substantial
uncertainties are caused  by the shift of $m_c$ and rather broad interval
of the inverse moment $\lambda_B$ of the $B$-meson DA.

To parameterize the charm-loop effect for
the $B\to K^* \ell^+\ell^-$ decay amplitude
we use its decomposition in the three invariant amplitudes
${\cal M}_i $   presented in Appendix~B
and the corresponding decompositions
(\ref{eq:Kstampl}) and (\ref{eq:HBKst}).
As a result
the terms proportional to $C_9$
in the amplitudes ${\cal M}_{1,2,3} $,
have to  be modified
in the following way:
\be
\Delta C_9^{(\bar{c}c,B\to K^*,\,{\cal M}_i)}(q^2)
=\left(C_1+3C_2 \right)g(m_c^2,q^2)
+2C_1\widetilde{g}\,^{(\bar{c}c,B\to K^*,\,{\cal M}_i)}(q^2)\,,
\label{eq:deltC9BKst}
\ee
where, respectively,
\begin{eqnarray}
\widetilde{g}\,^{(\bar{c}c,B\to K^*,\,{\cal M}_1)}(q^2) &=& -
{32\pi^2\over 3} { (m_B+m_{K^*})\widetilde{\V}_1(q^2) \over q^2
V^{BK^*}(q^2)}\,,
\nonumber\\
\widetilde{g}\,^{(\bar{c}c,B\to K^*,\,{\cal M}_2)}(q^2) &=& { 64
\pi^2 \over 3 } {(m_B- m_{K^{\ast})} \over q^2 }
\frac{\widetilde{\V}_2(q^2)}{A_1^{BK^*}(q^2)} \,,
\nonumber\\
\widetilde{g}\,^{(\bar{c}c,B\to K^*,\,{\cal M}_3)}(q^2) &=& { 64
\pi^2 \over 3  } \bigg [{ (m_B+m_K^*)\widetilde{\V}_2(q^2) \over
q^2 A_2^{BK^*}(q^2)}
\nonumber\\
&& + { \widetilde{\V}_3(q^2) \over (m_B- m_{K^{\ast}})
A_2^{BK^*}(q^2)} \bigg ] \, .
\end{eqnarray}
Note that the nonfactorizable contributions to
$\Delta C_9^{(\bar{c}c,B\to K^*,\,{\cal M}_i)}$ are enhanced
at small $q^2\geq 4m_{\ell}^2$ with respect to the factorizable ones
due to the virtual photon propagator (the same enhancement
as in the $C_7$ contribution).
Our predictions for $\Delta C_9^{(\bar{c}c,B\to K^*,\,{\cal M}_i)}(q^2)$
are plotted in Fig.~\ref{fig:BKstres}.
In Table~\ref{tab:gtilde}
we present a sample of numerical results and estimated uncertainties
for $\widetilde{g}\,^{(\bar{c}c,B\to K^*,\,{\cal M}_i)}(1 \mbox{GeV}^2)$.
The charm-loop effect and its soft-gluon part in $B\to K^*\ell^+\ell^-$
transitions is predicted to be considerably larger
than in $B\to K \ell^+\ell^-$, having also  a larger
uncertainty. Moreover, the
nonfactorizable contributions in $B\to K^*$ have the same sign as
the factorizable ones, which leads to an additional enhancement
of the charm-loop effect.

We come to an important conclusion that the calculated charm-loop
correction  \\ $\Delta C_9^{(\bar{c}c,B\to K)}(q^2)$ in $B\to K
\ell^+\ell^-$ remains small, not exceeding (within uncertainties)
$\sim5\%$ of the $C_9$ value at $0< q^2<$ 4.0 GeV$^2$. In  $B\to
K^* \ell^+\ell^-$  the same ratio may reach (adding up the
estimated uncertainties) as much as 20\% of $C_9$ at $1.0
<q^2<4.0$ GeV$^2$ and is inflated at smaller $q^2$. In both
decays, especially in $B\to K^* \ell^+\ell^-$, the nonfactorizable
soft-gluon part of $\Delta C_9$ plays a decisive role. At $q^2\geq
4.0$ GeV$^2$ where OPE for the charm-loop effect starts to diverge
we will use a phenomenological ansatz for $\Delta
C_9^{(\bar{c}c,B\to K^*,\,{\cal M}_i)}(q^2)$ presented below, in
Sect.~7.

\section{Charm-loop effect in $B\to K^*\gamma$}

We are in a position to predict the charm-loop effect
also in $B\to K^*\gamma$ which is simply a by-product of our calculation
for $B\to K^* \ell^+\ell^-$ at $q^2=0$. Importantly, in this case only the nonfactorizable
 soft-gluon emission contributes. With the effective
resummation of this effect,
our result goes beyond the estimates in \cite{KRSW} and \cite{BZJ}
where the matrix elements of the local quark-gluon
operator were calculated.

To specify the normalization, in Appendix B  we present
the dominant contribution to the $B \to K^{\ast} \gamma$  amplitude,
due to the operator $O_7$. The charm-loop effect
is included in this amplitude by the following additions to the Wilson coefficient
$C_7^{eff}$ multiplying the
first and second kinemetical structures in this
amplitude, respectively:
\be
C_7^{eff} \to C_7^{eff} + [\Delta C_7^{(\bar{c}c,B\to K^*\gamma)}]_{1,2}\,,
\ee
where
\ba
~~[\Delta C_7^{(\bar{c}c,B\to K^*\gamma)}]_{1}
=\frac{32 \pi^2}{3} \frac{C_1\tilde{\mathcal{V}}_1(0)}{
(m_b+m_s)T_1^{B K^{\ast}}(0)}\,,
\nonumber \\
~~
[\Delta C_7^{(\bar{c}c,B\to K^*\gamma)}]_{2}
=- \frac{64 \pi^2}{3} \frac{C_1\tilde{\mathcal{V}}_2(0)}{
(m_b-m_s)T_1^{B K^{\ast}}(0)} \,.
\label{eq:BKstg}
\ea

The numerical analysis reveals that the corrections to both
amplitudes are approximately  equal:
\begin{eqnarray}
\big[\Delta C_7^{(\bar{c}c,B\to K^*\gamma)}\big]_{1} \simeq
\big[\Delta C_7^{(\bar{c}c,B\to K^* \gamma)}\big]_{2} = (-1.2 ^{+0.9}_{-1.6}
)\times 10^{-2} \,,
\label{eq:deltaC7}
\end{eqnarray}
amounting up to 8 \% of the coefficient $C_7^{eff}$.
Our estimate yields larger magnitude (also with larger
uncertainty) than predicted in \cite{BZJ}
where the local OPE and LCSR with $K^*$-meson DA's were used.
Normalizing the hadronic matrix elements calculated
in \cite{BZJ} as in (\ref{eq:BKstg}) and using their input, we obtain
\begin{eqnarray}
~~[\Delta C_7^{(\bar{c}c,B\to K^*\gamma)}]_{1}^{BZ}= (-0.39 \pm 0.3)\times 10^{-2} \, ,
\nonumber\\~~
[\Delta C_7^{(\bar{c}c,B\to K^*\gamma)}]_{2}^{BZ}= (-0.65\pm 0.57) \times 10^{-2} \,.
\end{eqnarray}
Note that if we also use the local OPE limit, both amplitudes
in (\ref{eq:deltaC7}) have about 40\% larger magnitudes,
approaching the  estimates obtained
from the three-point QCD sum rules in \cite{KRSW}.

\section{Accessing large $q^2$ with dispersion relation }

Returning to the hadronic matrix element (\ref{eq:matr})
we use its analyticity in the variable $q^2$, expressed in the form of
dispersion relation.
Considering first the $B\to K $ case, we obtain the corresponding spectral density,
inserting in (\ref{eq:matr})
the full set of hadronic states with the  quantum numbers of $J/\psi$
between the $c$-quark current and the four-quark operator.
Introducing the hadronic matrix elements
\be
\langle 0 | \bar{c}\gamma^\rho c|\psi(q)\rangle = \epsilon_{\psi}^\rho m_\psi
f_\psi\,,
\label{eq:fpsi}
\ee
\be
\langle K(p)\psi(q) |C_1 O_1+
C_2 O_2|B(p+q)\rangle = (\epsilon_{\psi}^*\cdot p)m_\psi
A_{B\psi K }\,,
\label{eq:ABKpsi}
\ee
where $\psi=J/\psi,\psi(2S) $ and
$\epsilon_{\psi}$ is the polarization vector,
we obtain the dispersion relation for the invariant amplitude:
\ba
{\cal H}^{(B\to K)}(q^2) = {\cal H}^{(B\to K)} (0) + q^2\Big[
\sum_{\psi=J/\psi,\psi(2S)} \frac{f_\psi A_{B \psi K} }{m_\psi^2(m_\psi^2-q^2-
im_\psi\Gamma^{tot}_\psi)}
\nonumber\\
+\int_{4m_D^2}^{\infty} ds \frac{\rho(s)}{s(s-q^2-i\epsilon)}\Big]\,,
\label{eq:BKdisp}
\ea
where $\rho(s)$ is the spectral density
of $\psi$-resonances and continuum  $c\bar{c}$-states located
above the open charm threshold, $s\geq 4m_D^2$.
In (\ref{eq:BKdisp}) we include the small total widths of $J/\psi$
and $\psi(2S)$, but neglect the complex phase everywhere
beyond the immediate vicinity of the resonances.
One subtraction at $q^2=0$ takes into account that
the spectral density (\ref{eq:im1}) of the factorizable loop
contained in ${\cal H}^{(B\to K)}(q^2)$  tends to a constant
at $s\!\to\! \infty$.  Note that the soft-gluon
contribution has a convergent dispersion integral as follows
from the spectral density  (\ref{eq:im2}).
The parameters of $\psi$-poles in (\ref {eq:BKdisp})
are presented  in Appendix C, including
the decay constants $f_\psi$  and the absolute values of
the invariant amplitudes $A_{B \psi K}$ calculated from
the measured $B\to \psi K$ widths.

It is well known that the naive factorization approximation for
these amplitudes: $A_{B \psi K}=(C_2+C_1/3)f_\psi
f^+_{BK}(m_\psi^2) $ is not consistent with the experimental data,
indicating sizeable nonfactorizable corrections. Moreover, both
amplitudes $A_{B J/\psi K}$ and $A_{B\psi(2S) K}$ are expected to
have complex phases generated by the strong final-state
interactions in these nonleptonic decays \footnote{ These phases
originate from the discontinuities of the $B\to \psi K$ amplitudes
in the variable $(p+q)^2$ and are not related to the analytical
properties in the variable $q^2$. }. Note that the diagrams with
perturbative  gluons (e.g., the $O(\alpha_s)$ diagram in
Fig.~1c,d) generate complex phases in ${\cal H^{B\to K} }(q^2)$.
One can speculate that they are dual to the  final-state
interaction phases in $A_{B \psi K}$, in a certain analogy with
the QCD factorization approach \cite{BBNS}. The diagrams with
perturbative gluons are not included in our calculation of  ${\cal
H^{B\to K} }(q^2)$, hence we also neglect  complex  phases in the
hadronic part of the dispersion relation. On the other hand, we
notice that after taking  the soft-gluon emission into account,
the amplitude ${\cal H}^{(B\to K)}(q^2)$, is not positive-definite
anymore. Indeed, the nonfactorizable part $\tilde{\A}$ of this
amplitude obtained from LCSR has a negative sign with respect to
the factorizable part $\A$. Therefore, we relax the positivity
condition for the spectral density in (\ref{eq:BKdisp}), allowing,
e.g., different signs between the residues of $\psi$-poles
\footnote{The sign of the lowest $J/\psi$ contribution with
respect to the calculated amplitude in l.h.s. of (\ref{eq:BKdisp})
is simply fixed from the higher  derivatives of the dispersion
relation over $q^2$ at small $q^2$ where all other contributions
in r.h.s. are power suppressed.}.   This makes (\ref{eq:BKdisp})
essentially different from the dispersion relations used earlier
in \cite{KS,ABHH}.

The most complicated part of the dispersion relation
(\ref{eq:BKdisp}) is the spectral density $\rho(s)$ above the
open-charm threshold. It includes an interplay of broad charmonium
resonances and continuum states. Since we are working beyond the
factorization approximation, we do not attempt to describe this
spectral function as a sum over higher $\psi$ resonances and a
tail determined by quark-hadron duality with the $c$-quark loop as
in \cite{KS}. In future $\rho(s)$ can be partially determined
and/or constrained up to $s=(m_B-m_K)^2$, employing the
experimental data on the nonleptonic decays $B\to \psi K$,
($\psi=\psi(3770),...$) and $ B\to \bar{D}D K, \bar{D}^*D K, ...
$. Note also, that at $q^2$ approaching the upper threshold
$(m_B-m_K)^2$, the dispersion relation will also be influenced
by the singularities in $\rho(s)$ related to the intermediate
$b\bar{s}$ states. Hence we refrain from using the quark-hadron
duality approximation for the integral over $\rho(s)$ in
(\ref{eq:BKdisp}). The simplest possible choice is to model this
integral  by an effective pole: \be \int_{4m_D^2}^{\infty} ds
\frac{\rho(s)}{s(s-q^2)} \simeq \frac{a^{(B\to K)}}{m_*^2-q^2} \,.
\label{eq:rhoint} \ee The next step is to match the dispersion
relation to the function ${\cal H}^{(B\to K)}(q^2)$ obtained from
OPE and LCSR at $q^2\ll 4m_c^2$. More specifically, we use as an
input the l.h.s. of (\ref{eq:BKdisp}) calculated at $-4m_c^2\leq
q^2 \leq 2$ GeV$^2$, so that the subtraction constant ${\cal
H}^{B\to K} (0)$ is also determined. Provided the absolute values
of  $\psi$-poles in (\ref{eq:BKdisp}) are fixed by experimental
data \cite{HFAG}, this matching allows one to fit the parameters
of the effective pole. Importantly, the fit clearly favours a
negative relative sign of $J/\psi$ and $\psi(2S)$ contributions,
and yields for the central values of the input \be m_*=4.06 \mbox{
GeV} , ~~ a^{(B\to K)}= 0.06\times 10^{-3}\,. \label{eq:BKhfit}
\ee The fitted residue of the effective pole turns out to be much
smaller than residues of $J/\psi$ and $\psi(2S)$ in
(\ref{eq:BKdisp}): \be \frac{f_{J/\psi} A_{B J/\psi
K}}{m_{J/\psi}^2}= 1.34\times 10^{-3}\,,~~ \frac{f_{\psi(2S)} A_{B
\psi(2S) K}}{m_{\psi(2S)}^2}= -0.90\times 10^{-3}.
\label{eq:apsipsi2s} \ee To investigate the sensitivity of the
dispersion relation to the  adopted ansatz (\ref{eq:rhoint}), we
also used, as an alternative option, the conformal mapping $q^2\to
z$  and the $z$-expansion  for the integral  (\ref{eq:rhoint}),
making use of the fact that it has no singularities at
$q^2<4m_D^2$. The resulting dispersion representation for ${\cal
H}^{(B\to K)}(q^2)$ obtained after fitting the coefficients of $z$
expansion numerically differs very little from the one with the
effective pole, again  favoring the sign  pattern as in
(\ref{eq:apsipsi2s}), hence we adopt the latter as a default
model.

Finally, we express the dispersion representation of the charm-loop effect
in terms of the correction
$\Delta C_9^{(\bar{c}c,B\to K)}(q^2)$  defined as in (\ref{eq:deltC9}).
Our numerical prediction  is plotted in Fig.~\ref{fig:BKresfit}
up to $q^2=m_{\psi(2S)}^2$.
At $q^2\leq 4$ GeV$^2$ it coincides with the calculated
result shown in Fig.~\ref{fig:BKres}.
The dashed region in  Fig.~\ref{fig:BKresfit}
indicates now all uncertainties, in particular
the one  which is not related with our calculation and corresponds
to the substantial scale-variation
of the combination of Wilson coefficients in the factorizable part.
Note that at $q^2>$ 4  GeV$^2$, where we rely on the dispersion relation,
it is not possible to split $\Delta C_9^{(\bar{c} c, B \to K)}$
into factorizable  and nonfactorizable parts.
The predicted charm-loop effect in $B\to K\ell^+\ell^-$
is numerically unimportant at least up to $q^2\sim 5-6$ GeV$^2$.
Within uncertainties it is even consistent
with zero, due to possible cancellation of factorizable and
nonfactorizable contributions to $\Delta C_9(q^2)^{(\bar{c} c, B \to K)}$
at certain combinations of the input parameters.

Numerically, we find that the correction to $C_9$ is well reproduced
by the following simple parameterization valid at $0<q^2<9$ GeV$^2$
\begin{eqnarray}
\Delta C_9^{(\bar{c} c, B \to K)} (q^2)=
\frac{\Delta C_9^{(\bar{c} c, B \to K)} (0) +
r_1^{(B\to K)}\frac{q^2}{m_{J/\psi}^2 }}{
1- r_2^{(B\to K)} {q^2 \over m_{J/\psi}^2 }} \, ,
\label{eq:fitC9BK1}
\end{eqnarray}
with the calculated
\be
\Delta C_9^{(\bar{c} c, B \to K)}
(0)=0.17^{+0.09}_{-0.18} \ee and fitted  \be
r_1=0.01^{+0.13}_{-0.08}, ~~r_2=1.02^{+0.01}_{-0.01}.
\label{eq:dCBKfit}
\ee
This representation can be used in the
phenomenological analysis of $B\to K \ell^+\ell^-$.

\FIGURE[t]{
\includegraphics[scale=0.9]{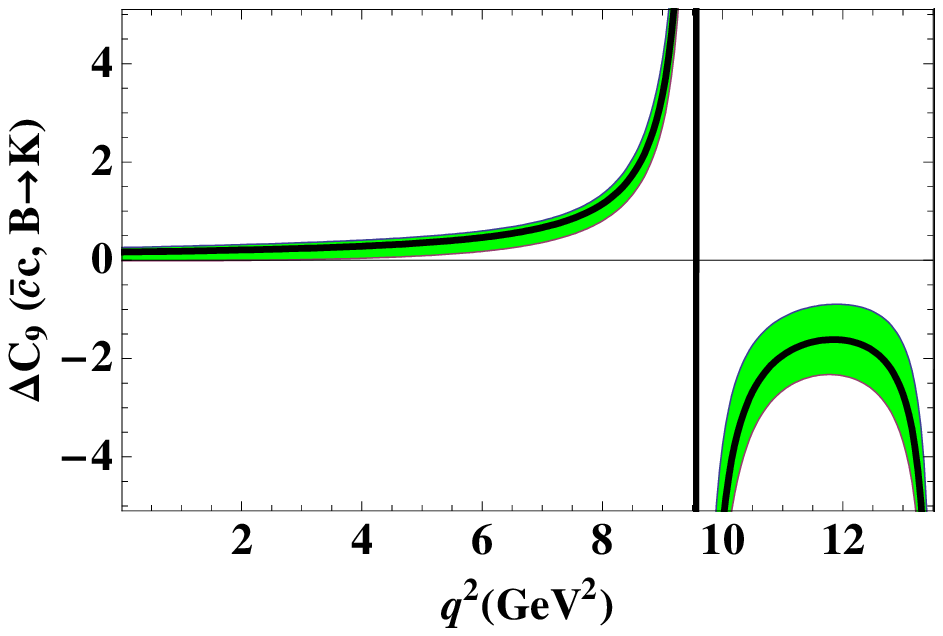}
\caption{\it The charm loop  contribution to the Wilson
coefficient $C_9$ for $\bar{B}_0 \to \bar{K} l^{+} l^{-}$ at $q^2$
below the open charm threshold, obtained from the dispersion
relation fitted to the OPE result at $q^2\ll 4m_c^2$. The central
values are denoted by dashed line, shaded area indicates the
estimated uncertainties.} \label{fig:BKresfit}
}
\FIGURE[t]{
\hspace{1.5 cm}\includegraphics[scale=0.7]{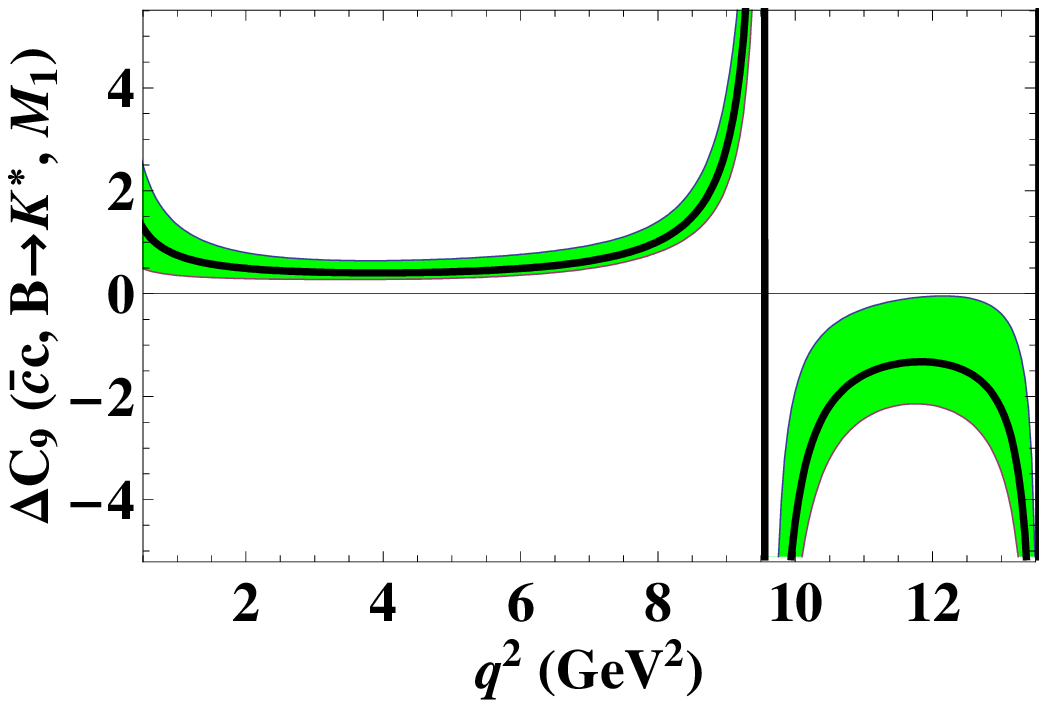}\\
\includegraphics[scale=0.7]{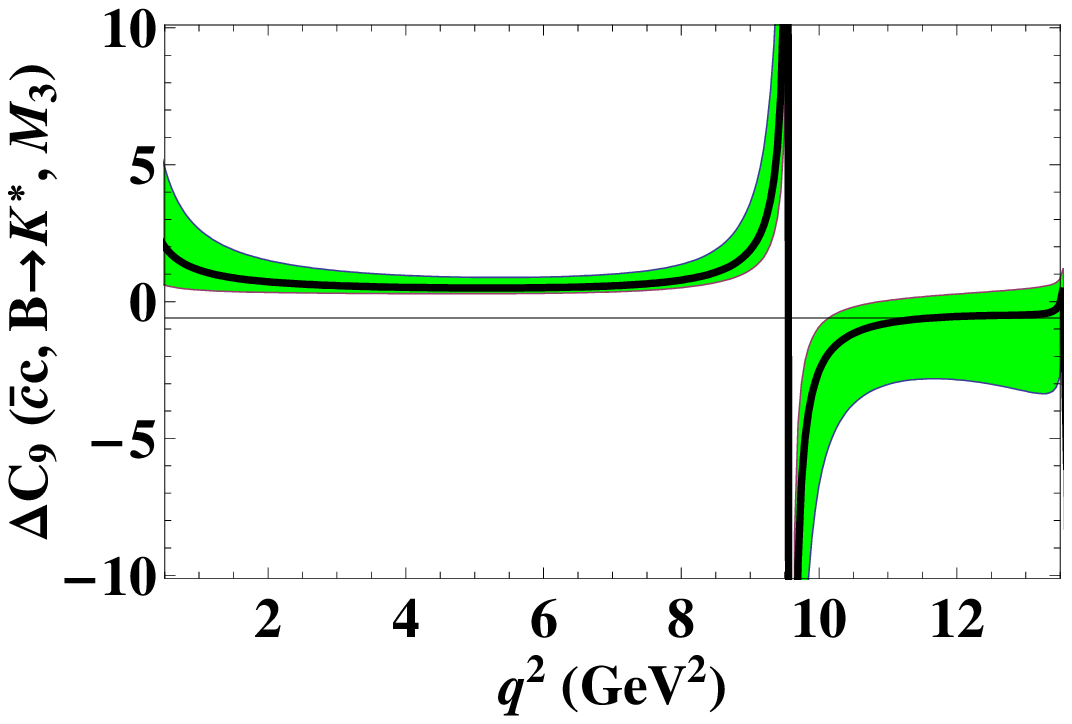}
\caption{The same as in Fig.~\ref{fig:BKresfit}
 for $\bar{B}_0 \to \bar{K^{\ast}} l^{+} l^{-}$. }
\label{fig:BKstdispC9}
}

The dispersion approach described above is extended to the
charm-loop effect in $B\to K^*\ell^+\ell^-$, in which case the
hadronic matrix element  (\ref{eq:matr}) consists of three
invariant amplitudes ${\cal H}_{i}^{(B\to K^\ast)}(q^2)$,
$i=1,2,3$. For the first two ones we employ dispersion relations
with two subtractions (taking into account one  extra power of
$q^2$ in the factorizable part):
\ba  {\cal H}_i^{(B\to
K^\ast)}(q^2) &=& {\cal H}_i^{(B\to K^\ast)}(0) + q^2\frac{d}{dq^2}{\cal H}_i^{(B\to K^\ast)}(0) \\
&&  + (q^2)^2 \bigg [ \sum_{\psi=J/\psi, \psi(2S)} \frac{
f_{\psi}A^{(i)}_{B\psi K^*}}{m_{\psi}^3 (m_{\psi}^2-q^2
-im_\psi\Gamma^{tot}_\psi)}  + h_i(q^2) \bigg ] \,,~~ (i=1,2)\,,
\label{eq:dispBKst1}  \nonumber
\ea
 whereas for the third amplitude a
combination of two dispersion relations has to be used yielding:
\ba {\cal H}_3^{(B\to K^\ast)}(q^2) = \sum_{\psi=J/\psi, \psi(2S)}
\frac{ f_{\psi}A^{(3)}_{B\psi K^*}}{m_{\psi} (m_{\psi}^2-q^2
-im_\psi\Gamma_\psi^{tot})}+ h_3(q^2)\, . \label{eq:dispBKst3} \ea
In the above, $A^{(i)}_{B\psi K^*}$ are the invariant amplitudes
determining  $B\to\psi K^*$ nonleptonic decays. They can be
expressed via transversity amplitudes (see Appendix C for the
definitions of the latter): \ba A^{(1)}_{B\psi K^*} &=& \frac{
\sqrt{2} A^\perp_{B\psi K^*}}{m_B^2 \lambda_{B\psi K^*}^{1/2}},~~
\hspace{5 cm} A^{(2)}_{B\psi K^*}= \frac{ -A^{\parallel}_{B\psi
K^*}}{ \sqrt{2} ( m_{B}^2-  m^2_{K^{\ast}})} \,,
\nonumber\\
A^{(3)}_{B\psi K^*} &=& \bigg \{ \Big[ 2 m_{K^{\ast}} m_{\psi}
A^0_{B\psi K^*} -\frac{A^{\parallel}_{B\psi K^*}}{\sqrt{2}} (m_B^2
- m_{\psi}^2-m_{K^{\ast}}^2) \Big] \frac{ m_B^2 - m_{K^{\ast}}^2
}{m_B^4 \lambda_{B\psi K^*}} + {A^{\parallel}_{B\psi K^*} \over
\sqrt{2} } \bigg \} \,. \hspace{1.0 cm} \ea The amplitudes
$A^{(\perp,\parallel,0)}_{B\psi K^*}$ and their relative phases
are extracted  using the kinematical analysis  of $B\to J/\psi
K^*$ and $B\to \psi(2S) K^*$ decays (see, e.g.,\cite{BRY}),
together with the latest data  on the angular distributions
\cite{BaBar}. The integrals over the spectral density of higher
states denoted as $h_i(q^2)$ in (\ref{eq:dispBKst1}) and
(\ref{eq:dispBKst3}) are parameterized with the help of effective
poles: \be h_i(q^2)= \frac{a_i^{(B\to K^*)}}{m_{*i}^2-q^2} \,,
\label{eq:hi} \ee After that the dispersion relations are fitted
to the calculated ${\cal H}_i^{(B\to K^\ast)}(q^2)$ at $q^2\ll
4m_c^2$. Without going into further details, we only mention that
in this case the pattern of relative signs and hierarchy of
contributions is very similar to the $B\to K$ case. In particular,
the position of the effective pole coincides within small
uncertainties with the one in (\ref{eq:BKhfit}).

With the help of the dispersion relation we finally obtain
the corrections $\Delta C_9^{(\bar{c} c, B \to K^{\ast},{\cal M}_i)}$
presented in Fig.~\ref{fig:BKstdispC9}.
\FIGURE[t]{
\includegraphics[scale=1.0]{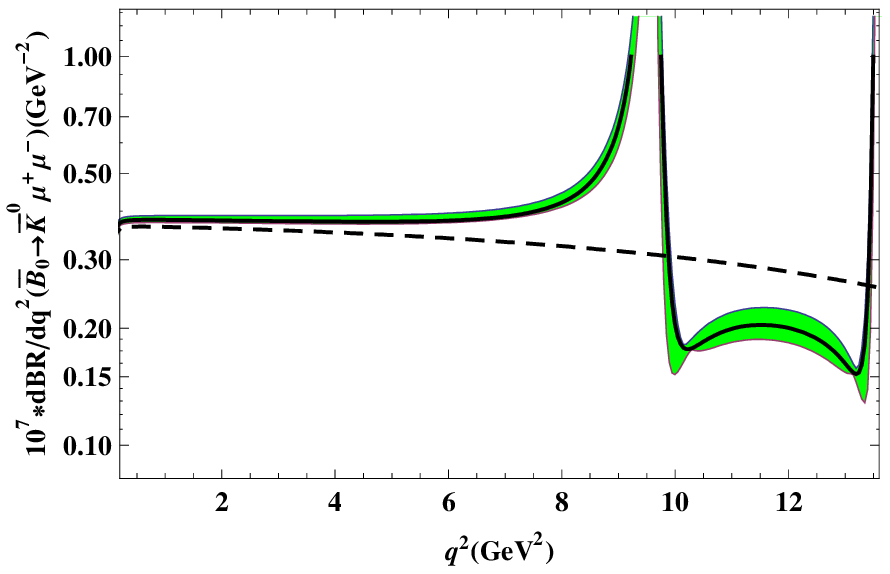}
\caption{\it The differential width of $\bar{B}_0 \to \bar{K}
\mu^{+} \mu^{-}$, including the charm-loop effect calculated with
the central values of input (solid, the shaded area indicates
estimated uncertainties) and without this effect (dashed).}
\label{fig:dGammaBK} }

\TABLE{
 \caption{\it Parameters of the ansatz
(\ref{eq:deltCBKstfit}) for $\Delta C_9^{(\bar{c} c, B \to
K^{\ast}, M_i)}(q^2)$ valid at $1.0 < q^2< 9.0$ {\rm GeV}$^2$.}
\begin{tabular}{|c|c|c|c|}
   \hline
&&&\\[-3mm]
parameters &value at $\bar{q}^2=1.0$ GeV$^2$ &$r_1^{(B\to K^*,{\cal
M}_i)}$&
$r_2^{(B\to K^*,{\cal M}_i)}$\\[1mm]
\hline&&&\\[-3mm]
  $\Delta C_9^{(\bar{c} c, B \to K^{\ast}, M_1)}$ & $0.72^{+0.57}_{-0.37}$  & $0.10^{+0.02}_{-0.00}$  &
$1.13^{+0.00}_{-0.01}$    \\[1mm]
   $\Delta C_9^{(\bar{c} c, B \to K^{\ast}, M_2)}$  & $0.76^{+0.70}_{-0.41}$  & $0.09^{+0.01}_{-0.00}$  &
$1.12^{+0.00}_{-0.01}$  \\[1mm]
   $\Delta C_9^{(\bar{c} c, B \to K^{\ast}, M_3)}$  & $1.11^{+1.14}_{-0.70}$  & $0.06^{+0.04}_{-0.10}$  &
$1.05^{+0.05}_{-0.04}$ \\[1mm]
   \hline
\end{tabular}
\label{tab:deltaC9}
}
For the phenomenologically interesting region
$1.0<q^2<9.0$ GeV$^2$ we suggest the following
numerical parameterizations of these corrections:
\begin{eqnarray}
\Delta C_9^{(\bar{c} c, B \to K^{\ast}, {\cal M}_i)} (q^2)=
\frac{r_1^{(B\to K^*,{\cal
M}_i)}\left(1-\frac{\bar{q}^2}{q^2}\right)+ \Delta C_9^{(\bar{c}
c, B \to K^{\ast}, {\cal M}_i)} (\bar{q}^2)
\frac{\bar{q}^2}{q^2}}{1+ r_2^{(B\to K^\ast,{\cal
M}_i)}\frac{\bar{q}^2-q^2}{m_{J/\psi}^2 }}\,,
\label{eq:deltCBKstfit}
\end{eqnarray}
where the calculated values of
$\Delta C_9^{(\bar{c} c, B \to K^{\ast}, {\cal M}_i)}(\bar{q}^2 = 1 {\rm GeV^2})$  and the fitted values of $r_{1,2}^{(B\to K^\ast,{\cal M}_i)}$
are collected in Table~\ref{tab:deltaC9}.

In the region between $J/\psi$ and $\psi(2S)$
our results for $\Delta C_9^{\bar{c} c, B \to K^{(\ast)}}(q^2)$
provide  at most crude estimates. Nevertheless, this region
is interesting because the predicted destructive interference
between the $J/\psi$ and $\psi(2S)$ poles manifests
itself in the form of a characteristic maximum located in the middle.
In case of the constructive interference (which is demanding an unnaturally
big, destructively interfering  contribution of higher $\psi$-states
in order to satisfy the dispersion relation) the maximum is replaced
by a monotonously increasing curve.

At $q^2>m_{\psi(2S)}^2$, the dispersion relation becomes too
complicated to be treated by any simple model, and the estimate of
the charm-loop effect remains an open problem. From what we
discussed above, it is obvious, that neither the approximation
$\{$ $\bar{c}c$-loop $\oplus$ gluon corrections$ \}$, nor a simple
sum over $\psi$ resonances can provide an adequate description of
this effect in  $B\to K^{(*)}\ell^+\ell^-$.

We are now in a position to investigate the impact
of the predicted charm-loop effect on the observables in
$B\to K^{(\ast)} \ell^+\ell^-$.
With the decay amplitudes defined in Appendix B
the differential widths
are calculated
adding the charm-loop corrections to $C_9$.
Since we are only interested in this effect,
the other small  nonfactorizable contributions
(e.g.. the loops due to the quark-penguin operators or
$u$-quark loops) are not taken into account.
Remember that our analysis also does not include
perturbative nonfactorizable effects, hence our
calculated widths are strictly speaking not yet the
complete predictions to be compared with the data.
Furthermore, for the calculation of the $B\to K\ell^+\ell^-$
differential width we use the $B\to K$
vector and tensor form factors obtained from more accurate LCSR
with kaon DA's \cite{DKMMO,DM} (see Appendix B).
The result is plotted in Fig.~\ref{fig:dGammaBK}. As expected,
the charm-loop effect becomes essential only if one approaches
the $J/\psi$ resonance region.

The same effect is significantly more pronounced in the differential width
of $B \to K^{\ast} \ell^{+} \ell^{-}$,
where in each $M_i$-part of the decay amplitude
one has to replace $C_9$ by $C_9 +\Delta C_9^{(\bar{c}c, B\to K^*,{\cal M}_i)}$.
The result
is presented in Fig. \ref{fig:dGammaBKst} where we normalize the width to
its value at $q^2=1 $ GeV$^2$ in order to diminish
the large uncertainties contributing to the width by $B\to K^*$ form factors.
The latter are calculated from LCSR with $B$ meson DA's (see Appendix B).

\FIGURE[t]{
\includegraphics[scale=0.8]{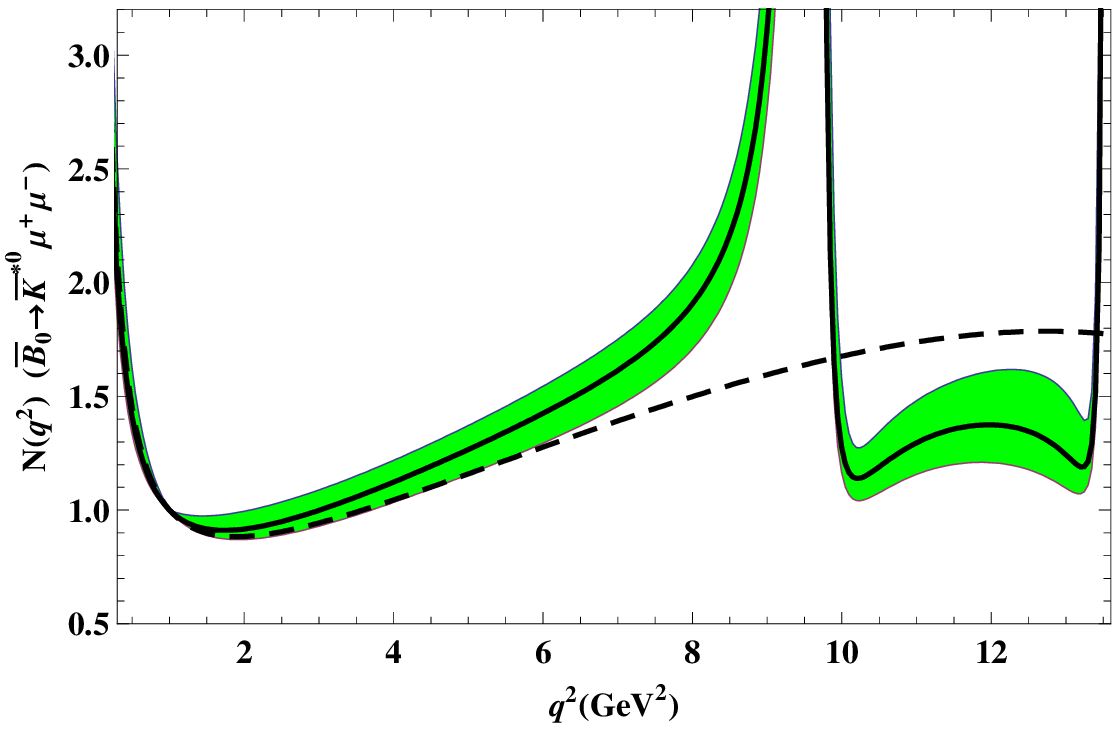}
\caption{\it The differential width of $\bar{B}_0 \to \bar{K}^\ast
\mu^{+} \mu^{-}$ normalized at $q^2=1.0$ {\rm GeV}$^2$, The
notations are the same as in Fig.~8. } \label{fig:dGammaBKst} }


Finally, we consider the forward-backward asymmetry in $B\to
K^{\ast} \ell^+\ell^-$ (for definition see e.g. \cite{ABHH} ). In
our notations, the zero-point of this asymmetry  is determined by
the following equation
\begin{eqnarray}
&& C_7^{eff}{m_B( m_b + m_s ) \over q_0^2} \bigg [
\bigg(1+\frac{m_{K^{\ast}}}{m_B}\bigg) { T_1^{B K^{\ast}}(q_0^2)
\over V^{B K^{\ast}}(q_0^2)} +
\bigg(1-\frac{m_{K^{\ast}}}{m_B}\bigg) { T_2^{B K^{\ast}}(q_0^2)
\over A_1^{B K^{\ast}}(q_0^2) } \bigg] \nonumber \\ && +  C_9
+\frac12 \bigg[ \Delta C_9^{(\bar{c}c,B\to K^*,{\cal M}_1)}(q_0^2)
+ \Delta C_9^{(\bar{c}c,B\to K^*,{\cal M}_2)}(q_0^2)\bigg] =0 \,.
\end{eqnarray}

The numerical result for the asymmetry is plotted
in Fig.~\ref{fig:BKstllFBA}.
Solving the above equation and taking into account the uncertainties
we obtain
\begin{eqnarray}
q^2_0 = 2.9^{+0.2}_{-0.3} {\rm GeV^2} \,
\label{eq:q2zero}
\end{eqnarray}
This is close to the prediction of \cite{ABHH}
and the one (at the leading-order) of \cite{BFS}.
Since the above interval at least
marginally belongs to the validity region of QCD calculation,
the estimate (\ref{eq:q2zero}) is practically independent of
the dispersion-relation ansatz. Hence, we can clarify
the influence of  the nonfactorizable contribution
on the position of the zero-point. Switching this contribution
off, we observe a small shift of the central value
towards $q^2_0 = 3.2 ~{\rm GeV^2}$. Note that,
according to \cite{BFS}, the addition of perturbative NLO corrections,
causes a significant shift of $q_0^2$ upwards.

\FIGURE[t]{
\includegraphics[scale=0.8]{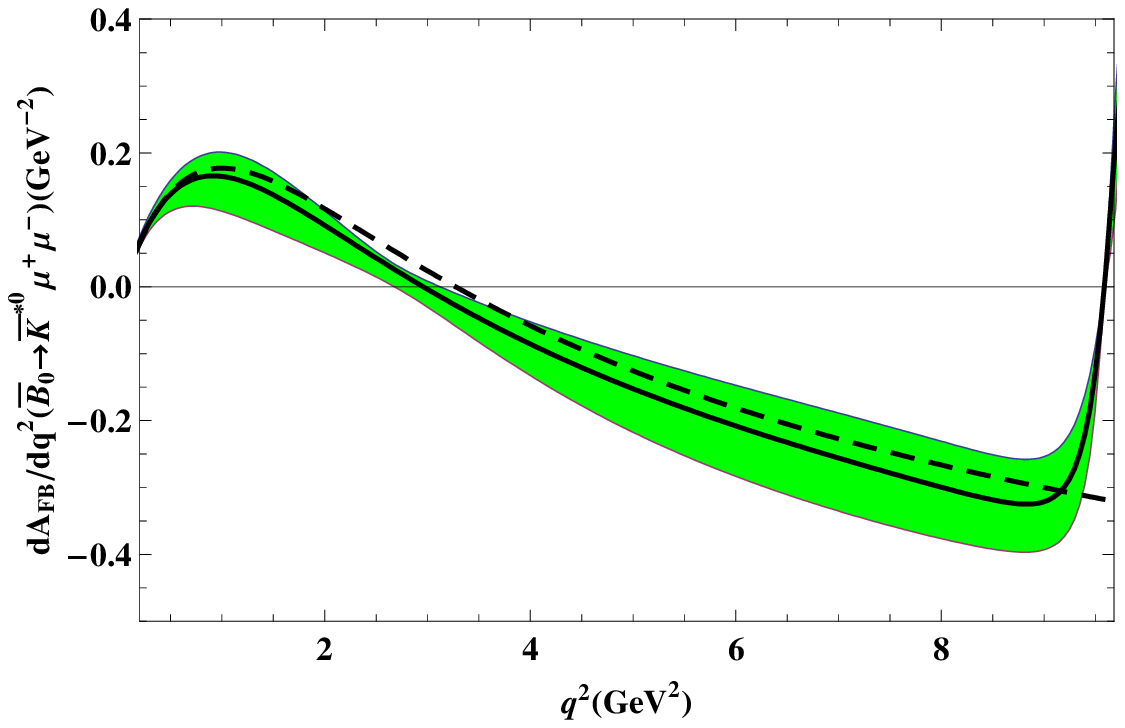}
\caption{\it The forward-backward asymmetry for $\bar{B}_0 \to
\bar{K^{\ast}} \mu^{+} \mu^{-}$ decay. The notations are the same
as in Figs 7,8.} \label{fig:BKstllFBA}
}

\section{Conclusion and outlook}

In this paper we went beyond the perturbative  approximation for
the charm-loop effect in  $B\to K^{(*)}\ell^+\ell^-$. In addition
to the leading-order $c$-quark  loop, the long-distance soft-gluon
emission which violates  factorization were taken into account.
Employing the OPE near the light-cone, we derived a  nonlocal
effective operator responsible for the soft-gluon emission from
the intermediate $c$-quarks. The $B\to K^{(*)}$ hadronic matrix
element of this operator has a structure of nonforward (flavour
changing) parton distribution, and has to be obtained from
nonperturbative QCD. We employed LCSR with $B$- meson DA's to
calculate these matrix elements. The same method and input was
used to obtain the $B\to K^{(*)}$ form factors entering the
factorizable contribution. In the decay amplitudes, the new
soft-gluon contribution becomes numerically important,  being
enhanced by its Wilson coefficient and by $\sim 1/q^2$ for  $B\to
K^{*}\ell^+\ell^-$ with respect to the factorizable part. The
characteristic power suppression of the soft-gluon contribution is
$\sim 1/(4m_c^2-q^2)$, signaling that the approximation
``perturbative loop $\oplus$ soft-gluon corrections'' is only
applicable at $q^2\ll 4m_c^2$.  At $q^2$ approaching the ${\bar
c}c$-threshold, multiple soft-gluon emission operators have to be
included, new hadronic  matrix elements proliferate and one
eventually looses control over OPE. A clear footprints of these
effects are the observed large violations of factorization in
$B\to \mbox{charmonium} + K^{(*)}$  decays.

LCSR with $B$-meson DA's used in this paper are not yet
sufficiently accurate. There are unaccounted $\sim 1/m_b$
corrections to the HQET correlation function, the gluon radiative
corrections are not taken into account and the parameters of
$B$-meson DA's, e.g., the inverse moment, still have large
uncertainties. The imperfection of sum rules eventually converts
into relatively large theoretical errors of our calculation.
Nevertheless, since we are investigating a small effect, at least
at low $q^2$, the achieved accuracy is reasonable. We believe that
in future also other methods, first of all, lattice QCD can be
used to calculate the $B\to K^{(*)}$ hadronic matrix elements
emerging from the light-cone OPE. Also the conventional LCSR with
light-mesons DA's and $B$ -meson interpolating current can be
employed for this purpose, with necessary modifications
\cite{AKBpipi}  (an artificial 4-momentum in the vertex of the
effective operator).

In this paper we also applied  the OPE-constrained dispersion relations
for the $B\to K^{(*)}$ hadronic amplitudes
with charm loop in order to
to access the large $q^2$ region, including charmonium resonances.
Our approach goes beyond the earlier ansatz \cite{KS,ABHH}
in terms of dispersion relation, since we include nonfactorizable
corrections  and use the  QCD result
at small and spacelike $q^2$ to constrain the integral over higher states.
This analysis clearly indicates a nontrivial interference between the
contributions of $J/\psi$ and $\psi(2S)$ states.

One conclusion of our study sounds rather pessimistic.
In our opinion, it is difficult,
if not impossible at all to make a reliable  prediction for the charm-loop
effect above $\psi(2S)$, based on QCD.
Although the actual effect could be small in this region, it will depend
on the interference of many charmonium states and
cannot be reliably constrained by OPE. In principle,
new accurate data on the branching fractions $B\to \psi K^{(*)}$ for $\psi$ states heavier than  $\psi(2S)$ and eventually also on $B\to \bar{D}D K^{(*)}$
can be used to saturate the spectral density of the
charm loop amplitude at large $q^2$.

In this paper we considered only one, albeit
important effect of four-quark operators with $c$ quarks
in $B\to K^{(*)}\ell^+\ell^-$.
Similar effects of quark loops with different flavours
stemming from CKM suppressed and/or quark-penguin operators
can be taken into account along the same lines. Note that
in the case of light-quark loops,
e.g., the $u$-quark loops important in $B\to \rho (\pi) \ell^+\ell^-$,
one has to expand at large spacelike $q^2$.
A separate interesting question deserving future study
is the potential role of soft
gluons in the weak annihilation contributions in rare
semileptonic $B$ decays.

\section* {Acknowledgments}
This work was supported by the Deutsche Forschungsgemeinschaft
under the  contract No. KH205/1-2. A.K. and T.M. are grateful to
the Galileo Galilei Insitute for Theoretical Physics for warm
hospitality during their visit, when the  important part of this
work was done. A.A.P. acknowledges the partial support by the RFFI
grant 08-01-00686. We are grateful to  P.~Ball, M.~Beneke,
M.~Gorbahn, Z.~Ligeti, M.~Misiak, and R.~Zwicky for useful
discussions and comments.

\section*{Appendix A: Effective Hamiltonian}
For convenience, here we list all those operators $O_i$ and their
Wilson coefficients $C_i$, entering $H_{eff}$ in (\ref{eq:Heff})
which are used throughout this paper omitting $O_{3-6}$ with very
small Wilson coefficients and $O_{8g}$ :
\begin{eqnarray}
 O_1 &=&
\left(\bar{s}_L\gamma_\rho c_L\right) \left(\bar{c}_L\gamma^\rho
b_L\right)\,,
~~~ O_2 =
\left(\bar{s}^j_{L}\gamma_\rho c^i_{L}\right) \left(\bar{c}^i_{L}\gamma^\rho
b^j_{L}\right)\,, \nonumber
\\
O_9 &=& \frac{\alpha_{em}}{4\pi}\left(\bar{s}_L\gamma_\rho
b_L\right) \left( \bar{l}\gamma^{\rho}l \right)\,, ~~~
O_{10}=\frac{\alpha_{em}}{4\pi}\left(\bar{s}_L\gamma_\rho
b_L\right) \left(\bar{l}\gamma^{\rho}\gamma_5 l \right)\, ,
\nonumber \\
O_{7\gamma}&=& -{e \over 16 \pi^2} \bar{s} \sigma_{\mu \nu} (m_s
L+ m_b R)b F^{\mu \nu }\,, \nonumber \label{eq:listoper}
\end{eqnarray}
where the notations $q_{L(R)}=\frac{1-(+)\gamma_5}{2}q$ and
$L(R)=\frac{1-(+)\gamma_5}{2}$ are used.
 We use the standard conventions for the operators
$O_i$, except the labelling of $O_1$ and $O_2$ is interchanged.

The sign convention for $O_{7 \gamma}$ and $O_{8 g}$ corresponds
to the covariant derivative $iD_\mu= i\partial_\mu+eQ_f A_{\mu} + g T^a
A^a_{\mu}$, where $Q_f$ is the fermion charge. In addition, the convention for
the Levi-Civita tensor adopted in this work is
${\rm Tr}\{\gamma^\mu\gamma^\nu\gamma^\rho\gamma^\lambda\gamma^5\}=4i
\epsilon^{\mu\nu\rho\lambda} \,,~ \epsilon^{0123}=-1.$

We use the Wilson coefficients $C_i$ calculated in the leading
approximation. They are given in Table \ref{tab:Wilsc}.
\TABLE[p]{ \caption{Numerical values of the  Wilson coefficients
at three different scales.}
\begin{tabular}{|c|c|c|c|c|}
  \hline
  $\mu$ (GeV)                 &  $0.5 m_b$    &  $ m_b$       & $1.5 m_b$          \\  \hline
   $C_1$                      &  $1.180$       &  $ 1.117$      & $ 1.090$         \\  \hline
   $C_2$                      &  $-0.380$       &  $-0.267$      & $ -0.214$       \\  \hline
$C_{7}^{eff}$             &  $-0.363$       &  $-0.319$      & $-0.298$ \\
\hline
   $C_9$                      &  $4.435$      &  $4.228$     & $4.034$      \\  \hline
   $C_{10}$                   &  $-4.410$      &  $-4.410$      & $-4.410$        \\  \hline
\end{tabular}
\label{tab:Wilsc}
}

\section*{Appendix B: Decay amplitudes
and form factors}


\subsection*{1. $B \to K \ell^{+} \ell^{-}$}
The dominant contributions to
the amplitude $B \to K \ell^{+} \ell^{-}$ stem from
the hadronic matrix elements of  the operators
$O_{7\gamma}$, $O_9$ and $O_{10}$:
\begin{eqnarray}
A (B \to K \ell^{+} \ell^{-}) &=& { G_F \over  \sqrt{2} }
{\alpha_{em} \over \pi} V_{tb} V_{ts}^{\ast} \Bigg[
\bar{\ell}\gamma_{\mu} \ell\, p^\mu\bigg( C_9 f^{+}_{BK}(q^2)
\nonumber \\
&& + {2 (m_b+m_s) \over m_B+m_K} C_7^{eff} f^{T}_{BK}(q^2) \bigg)
+ \bar{\ell} \gamma_{\mu} \gamma_5 \ell \, p^\mu C_{10}
f^{+}_{BK}(q^2) \bigg] \,, \hspace{2.7 cm}\label{decay amplitudes}
\end{eqnarray}
where we omit the contributions yielding
the lepton mass (the latter are included in the numerical analysis).
The tensor $B\to K$ form factor is defined as
\be
\langle K(p)|\bar{s} \sigma_{\mu \rho} q^{\rho}b|B(p+q)\rangle =
\bigg[q^2(2p_\mu +q_\mu) -(m_B^2-m_K^2)q_\mu\bigg]
\frac{i f^{T}_{BK}(q^2)}{m_B+m_K}\,.
\ee
\subsection*{ 2. $B \to K^* \ell^{+} \ell^{-}$}
For $B\to K^* \ell^+\ell^-$ decay amplitude we use the following
expression:
\begin{eqnarray}
A (B \to K^{\ast} l^{+} l^{-}) &=& { G_F \over 2 \sqrt{2} }
{\alpha_{em} \over \pi} V_{tb} V_{ts}^{\ast} \Bigg\{ \bar{l}
\gamma^{\mu} l \Big[\epsilon_{ \mu \nu \rho \sigma }
\epsilon^{\ast \nu} q^{\rho} p^{\sigma} {\cal M}_1(q^2)
\nonumber\\
&& -i \epsilon^{\ast}_{\mu} {\cal M}_2(q^2) + i  (\epsilon^{\ast}
\cdot q)p_{\mu}{\cal M}_3(q^2)\Big]
\\
&& +\bar{l} \gamma^{\mu} \gamma_5 l\Big[\epsilon_{ \mu \nu \rho
\sigma } \epsilon^{\ast \nu} q^{\rho} p^{\sigma} {\cal N}_1(q^2)
-i \epsilon^{\ast}_{\mu} {\cal N}_2(q^2)
+ i  (\epsilon^{\ast} \cdot q)p_{\mu}{\cal N}_3(q^2) \Big] \Bigg\}
\,, \hspace{1 cm}
\nonumber
\label{eq:ABKstll}
\end{eqnarray}
where
\begin{eqnarray}
{\cal M}_1(q^2) &=& C_9 { 2 V^{B K^{\ast}}(q^2)\over
m_B+m_{K^{\ast}}} + 4 C_7^{eff} { m_b + m_s \over q^2} T_1^{B
K^{\ast}}(q^2)\,,
\nonumber \\
{\cal M}_2(q^2)&=& C_9 (m_B+m_{K^{\ast}}) A_1^{B K^{\ast}}(q^2)
\nonumber \\ && +  2 C_7^{eff} (m_B^2 - m_{K^{\ast}}^2) { m_b +
m_s \over q^2} T_2^{B K^{\ast}}(q^2)\,,
\nonumber\\
{\cal M}_3(q^2) &=&
2 C_9 { A_2^{B
K^{\ast}}(q^2)  \over m_B+m_{K^{\ast}} }
\nonumber\\
&& + 4 C_7^{eff} { m_b - m_s \over q^2} \bigg(T_2^{B
K^{\ast}}(q^2) + {q^2 \over m_B^2 - m_{K^{\ast}}^2}
T_3^{BK^{\ast}}(q^2) \bigg) \,, \, \,\, \,\,\,\,  \label{eq:Mi}
\end{eqnarray}
and
\begin{eqnarray}
{\cal N}_1(q^2) &=& 2 C_{10} {V^{B K^{\ast}}(q^2) \over m_B+
m_{K^{\ast}}},\qquad {\cal N}_2(q^2)= C_{10} ({m_B +
m_{K^{\ast}}}) A_1^{B K^{\ast}}(q^2)\,,
\nonumber \\
{\cal N}_3(q^2)&=&
 2 C_{10}  { A_2^{B K^{\ast}}(q^2) \over m_B +
m_{K^{\ast}}}\, .
\end{eqnarray}
The tensor $B \to K^{\ast}$ form factors are defined as
\begin{eqnarray}
\langle K^{\ast}(p)|\bar{s} \sigma_{\mu \rho} q^{\rho} (1+
\gamma_5) b|B(p+q)\rangle = 2 i \epsilon_{\mu \nu \rho \sigma }
\epsilon^{\ast \nu } q^{\rho}  p^{\sigma}
T_1^{BK^{\ast}}(q^2)
\nonumber \\
+ [(m_B^2- m^2_{K^{\ast}}){\epsilon}^{\ast}_\mu
-(\epsilon^\ast \!\cdot q) (2p+q)_{\mu}] T_2^{B K^{\ast}}(q^2)
\nonumber \\
+  (\epsilon^\ast\! \cdot q) \bigg[q_{\mu} - {q^2 \over m_B^2-
m^2_{K^{\ast}}} (2 p+q)_{\mu}\bigg] T_3^{B K^{\ast}}(q^2) \, ,
\label{eq:BKsttf}
\end{eqnarray}
with $T_1^{B K^{\ast}}(0)=T_2^{B K^{\ast}}(0)$.

\subsection*{3. $B \to K ^* \gamma $}
The amplitude of $B \to K^{\ast} \gamma$  decay is
\begin{eqnarray}
A (B \to K^{\ast}\gamma) = { G_F \over  \sqrt{2} } {e \over 2 \pi^2 }
V_{tb}V_{ts}^{*} C_7^{eff}\Big\{
-(m_b+m_s) \epsilon_{ \mu \nu \rho \sigma }
\epsilon^{\ast \mu}_{\gamma} \epsilon^{\ast \nu}
q^{\rho} p^{\sigma}
\nonumber \\
+i (m_b- m_s) \big[(\epsilon^{\ast} \cdot
\epsilon^{\ast}_{\gamma}) (p \cdot q ) -
(\epsilon^{\ast} \cdot q)(\epsilon^{\ast}_{\gamma}
\cdot p)   \big]\Big \}T_1^{B K^{\ast}}(0)  \, ,
\label{eq:BKgamma}
\end{eqnarray}
where $\epsilon^{\ast}_{\gamma}$ is the polarization vector of the photon.
\subsection*{4. $B \to K ^{(*)} $ form factors }
The  $B \to K^{(\ast)}$ form factors used in the
factorizable parts of the charm-loop effect
calculation are substituted by the corresponding LCSR
with $B$ meson DA's. For the differential widths
we need an explicit numerical parameterization
of these and also tensor form factors.
We parameterize the $q^2$-dependence of all $B \to K^{(\ast)}$ form
factors with the $z$-parameterization similar to the one
suggested in \cite{BCL}, with one slope parameter:
\begin{eqnarray}
F(q^2) =\frac{ F(0)}{ 1- q^2 / m^2_{B_s(J^P)}} \bigg \{   1 +
b_1\bigg( z(q^2, t_0) - z(0, t_0) +\frac12\big[ z(q^2, t_0)^2
-z(0, t_0)^2\big ] \bigg ) \bigg \} \, , \label{eq:ffparam}
\end{eqnarray}
and
\begin{eqnarray}
&& z(q^2, \tau_0) = { \sqrt{\tau_{+}-q^2}-\sqrt{\tau_{+}-\tau_{0}} \over
\sqrt{\tau_{+}-q^2}+\sqrt{\tau_{+}-\tau_{0}}} \, , \nonumber \\
&& \tau_{+} = (m_B+m_{K^{(\ast)}})^2   \, ,
\qquad  \tau_{-}= (m_B-m_{K^{(\ast)}})^2  \nonumber \\
&& \tau_0 = \tau_{+} -\sqrt{\tau_{+}-\tau_{-}}\sqrt{\tau_{+}}\, .
\end{eqnarray}
The pole corresponds to the  $\bar{s}b$- resonance with
appropriate $J^P$. The $B\to K$ form factors are calculated from
more accurate LCSR with kaon DA's \cite{DKMMO,DM} at $q^2<12$
GeV$^2$ including the spacelike region, and the result is fitted
to the above parameterization. The $B\to K^*$ form factors are
calculated from LCSR with $B$ meson DA's \cite{KMO2} in the same
region. For the two tensor form factors $T_{2,3}$ the results are
new. The form factor values at $q^2=0$ and the fitted slope
parameters  are presented in Table \ref{B to K form factors},
where the masses of the $B_s$ states are \cite{PDG}:
$m_{B_s(0^-)}=5.366$ GeV, $m_{B_s^*(1^-)}=5.412$ GeV and
~$m_{B_s^*(1^+)}=5.829$ GeV. Within uncertainties, the form
factors obtaned from the two different LCSR agree with each other
(see\cite{KMO2}).
\TABLE[b]{
\begin{tabular}{|c|c|c|c|c|}
\hline   &&&&\\[-2mm]
form factor & $F^{i}_{B K^{(\ast)}}(0)$ & $b_1^{i}$ & $B_s(J^P)$&input
\\[1mm]
&&&&at $q^2<12$ GeV$^2$\\
\hline
&&&&\\[-2mm]
  $f_{BK}^{+}$ & $0.34 ^{+0.05}_{-0.02}$  & $-2.1^{+0.9}_{-1.6}$ &
$B_s^*(1^-)$&\\[1mm]
  $f_{BK}^{0}$ & $0.34 ^{+0.05}_{-0.02}$ & $-4.3^{+0.8}_{-0.9}$ & no pole
& LCSR \\[1mm]
  $f_{BK}^{T}$ & $0.39 ^{+0.05}_{-0.03}$ & $-2.2^{+1.0}_{-2.00}$ &
$B_s^*(1^-)$ &with $K$ DA's\\[1mm]
  \hline &&&&\\[-2mm]
  $V^{B K^{\ast}}$ & $0.36^{+0.23}_{-0.12}$& $-4.8^{+0.8}_{-0.4}$
&$B_s^*(1^-)$   &\\[1mm]
  $A_1^{B K^{\ast}}$ & $0.25^{+0.16}_{-0.10}$  & $0.34^{+0.86}_{-0.80}$ &
$B_s(1^+)$ & \\[1mm]
  $A_2^{B K^{\ast}}$& $0.23^{+0.19}_{-0.10}$ & $-0.85^{+2.88}_{-1.35}$ &
$B_s(1^+)$ & LCSR\\[1mm]
  $A_0^{B  K^{\ast}}$ & $0.29^{+0.10}_{-0.07}$  & $-18.2^{+1.3}_{-3.0}$  &
$B_s(0^-)$ &
with $B$ DA's\\[1mm]
  $T_1^{B  K^{\ast}}$ & $0.31^{+0.18}_{-0.10}$ & $-4.6^{+0.81}_{-0.41}$  &
$B_s^*(1^-)$ &\\ [1mm]
  $T_2^{B  K^{\ast}}$& $0.31^{+0.18}_{-0.10}$ & $-3.2^{+2.1}_{-2.2}$ &
$B_s(1^+)$&\\ [1mm]
  $T_3^{B  K^{\ast}}$ & $0.22^{+0.17}_{-0.10}$ & $-10.3^{+2.5}_{-3.1}$ &
$B_s(1^+)$ &\\[1mm] \hline
\end{tabular}\\[1mm]
\caption{The $B \to  K^{(\ast)}$ form factors from LCSR and their
$z$-parameterization.}
 \label{B to K form factors}}

\section*{Appendix C: Parameters of $\psi=\{J/\psi,\psi(2S)\}$
and  $B\to \psi K^{(*)}$ amplitudes} The necessary data on the two
lowest charmonium levels are collected in Table \ref{charmoniu
data}, where also the absolute values of the $B\to \psi K$ decay
amplitudes are given, calculated from: \be |A_{B \psi K }| =
\Bigg(\frac{8\pi BR(B\to \psi K)}{\tau(B) G_F^2 |V_{cs}|^2
|V_{cb}|^2 m_B^3 \lambda ^{3/2}_{B \psi K}}\Bigg)^{1/2}\,, \ee
where  $\lambda_{B \psi K}=\lambda(1,m_K^2/m_B^2,m_\psi^2/m_B^2)$,
and $\lambda(a,b,c)\equiv a^2+b^2+c^2-2 ab -2 ac -2bc$.

For $ B\to \psi K^* $
we use the decomposition in  transversity amplitudes:
\begin{eqnarray}
&& \langle K^{\ast}(p)\psi(q) | \left(C_1O_1+C_2 O_2\right)|
B(p+q) \rangle
\nonumber\\
&& = i \sqrt{2}{\epsilon}_{\alpha} {\epsilon_{\psi}}_{\beta} \Bigg
\{i\epsilon^{\alpha \beta \rho \tau} p_{\rho} q_{\tau}
\frac{A^\perp_{B\psi K^*}}{m_B^2 \lambda_{B\psi K^*}^{1/2}}
+g^{\alpha \beta }\frac{A^{\parallel}_{B\psi K^*}}{2}
\\
&& + \frac{1}{m_B^4 \lambda_{B\psi K^*}}(p+q)^{\alpha}
(p+q)^{\beta}\Big[ 2 m_{K^{\ast}} m_{\psi} A^0_{B\psi K^*}
-(m_B^2 - m_{\psi}^2-m_{K^{\ast}}^2)A^{\parallel}_{B\psi K^*}
\Big]  \Bigg\}\,, \hspace{1 cm} \nonumber
\end{eqnarray}
where $\lambda_{B\psi K^*}=\lambda(1,{ m^2_{K^{\ast}}/m_B^2},{ m_{\psi}^2/
m_B^2}) $, so that the decay  width is
\begin{equation}
\Gamma(B \to \psi K^{\ast})
=\frac{\lambda_{B\psi K^*}^{1/2}}{16\pi m_{B} } \bigg({4 G_F \over
\sqrt{2}} \bigg)^2 |V_{cb} V^{\ast}_{cs}|^2
\sum_{i=0,\parallel,\perp} |A^i_{B\psi K^*}|^2\,. \label{dr1}
\end{equation}

The polarization fractions $f_i(i=0,\parallel,\perp)$ are defined
as follows:
\begin{eqnarray}
f_i=\frac{|A_{B\psi K^*}^i|^2}{|A_{B\psi K^*}^0|^2+|A_{B\psi K^*}^\parallel|^2+
|A_{B\psi K^*}^\perp|^2}\;.
\end{eqnarray}

The transversity amplitudes in this decay can  be  determined from
the data on the branching fractions and polarization fractions
collected in  Table \ref{charmoniu data}. In addition, one can
determine the relative sign  between $A_0$ and $A_\parallel$.
\TABLE[b]{
\begin{tabular}{|c|c|c|}
\hline
\hline
state & $J/\psi$ & $\psi(2S)$\\
\hline
Mass ($ {\rm MeV}$)& $3096.916 \pm 0.011$ & $3686.09 \pm 0.04$  \\
$\Gamma_{tot} \,\, ({\rm keV}) $ & $93.2 \pm 2.1$ & $317 \pm 9$ \\
$\Gamma_{ll} \,\, ({\rm keV})$ & $5.55 \pm 0.14 \pm 0.02$ & $2.38 \pm 0.04$
\\
$f_{\psi} \,\,({\rm MeV})$ & $416^{+5}_{-6}$ & $297^{+3}_{-2}$ \\
$BR (\bar{B}^0 \to J/\psi K^{0})$ & $8.63 \pm 0.35  \times 10^{-4}$ & $6.55
\pm 0.66  \times 10^{-4}$ \\
$|A_{B \psi K}| (\rm MeV)$ & $31^{+3}_{-1}$  &
$41^{+4}_{-2}$ \\
$BR (\bar{B}^0 \to J/\psi K^{\ast 0})$  & $13.3 \pm 0.7 \times 10^{-4}$  &
$7.10 \pm 0.62 \times 10^{-4}$ \\
$f_0$ & $0.556 \pm 0.009 \pm 0.010$  & $0.48 \pm 0.05 \pm 0.02$ \\
$f_{\|}$& $0.211 \pm 0.010 \pm 0.006$  & $0.22 \pm 0.06 \pm 0.02$  \\
$f_{\perp}$ &  $0.233 \pm 0.010 \pm 0.005$ & $0.30 \pm 0.06 \pm 0.02$ \\
$|A_0 | (\rm GeV^3)$ & $0.27^{+0.00}_{-0.01}$ & $0.22^{+0.01}_{-0.02}$  \\
$|A_{\|}|  (\rm GeV^3)$ & $0.16^{+0.01}_{-0.00}$ & $0.15^{+0.02}_{-0.03}$
\\
$|A_{\perp } | (\rm GeV^3)$ &  $0.17^{+0.01}_{-0.00}$&
$0.17^{+0.02}_{-0.02}$ \\
\hline
\hline
\end{tabular}
\label{charmoniu data} \caption{The characteristics of $J/\psi$
and $\psi(2S)$ taken from \cite{HFAG,BaBar}. } }

\section*{Appendix D: The coefficients in the LCSR }
The resulting expressions for the coefficients
$C^{(F)}(q^2,u,\sigma,\omega,t)$ entering the sum rule
(\ref{eq:LCSRBK}), $F=\Psi_V, \Psi_{AV},...$ are presented here in a
form of the quadratic polynomials in the variable $\omega$,
$$
C^{(F)}(q^2,u,\sigma,\omega,t)=\sum\limits_{r=0,1,2}
C^{(F)}_r(q^2,u,\sigma,t)~\left(\frac{\omega}{m_B}\right)^r\,.
$$
The nonvanishing coefficients of these polynomials are:
\ba C^{(\Psi_V)}_0&=&(1-t) t \left[2 \text{q2} ((2 \text{sig}-5)
u+2)-2 \text{mB}^2
    (\text{sig}-1)^2 (u-1)\right]\nonumber \\
&& -\frac{1}{2} \text{q2} (2 \text{sig}-3) (2u-1)\,,
\nonumber \\
C^{(\Psi_V)}_1&=&4 (1-t) t u \left[\text{mB}^2 (\text{sig} (2
(\text{sig}-4)
    u-\text{sig}+5)+7 u-4)+2 \text{q2} (u-1)\right]
\nonumber \\
&& +\text{mB}^2 (2
    \text{sig}-3) u (2 u-1)\,,
\nonumber \\
C^{(\Psi_V)}_2&=&8 \text{mB}^2 (1-t) t u^2 (\text{sig} (4 u-3)-3
u+3)\,; \ea
\ba C^{(\Psi_{AV})}_0 &=&(1-t) t \big[\text{mB}^2 (\text{sig}-1)^2
(u-2)
 -\text{q2} (6 \text{sig}u+u-2)\big]+\text{q2} \text{sig} (2
u-1)\,,
\nonumber\\
C^{(\Psi_{AV})}_1 &=&
 -\frac{2 (1-t) t u }{\text{sig}-1} \bigg(\text{mB}^2
    (\text{sig}-1) \big[\text{sig} (2 \text{sig} (u-1)-7 u+4)+u-2\big]
\nonumber\\
&& +\text{q2} (2 \text{sig}+u-2)\bigg ) + 2 \text{mB}^2 \text{sig}
(1-2 u) u \,; \ea
\ba C^{(XA)}_0 &=& \frac{(1-t) t \left[\text{mB}^2
(\text{sig}-1)^2 (u-2)+\text{q2} ((5-6
    \text{sig}) u-2)\right]}{\text{mB} (\text{sig}-1)}
\nonumber\\
&& +\frac{\text{q2} (2
    \text{sig}-1) (2 u-1)}{2 \text{mB} (\text{sig}-1)}\,,
\nonumber\\
C^{(X_A)}_1 &=& \frac{4 (1-t) t u \left[\text{mB}^2
\left((\text{sig}-1)^2- ((\text{sig}-4)\text{sig}+2)
u\right)+\text{q2} (1-u)\right]}{\text{mB} (\text{sig}-1)}
\nonumber\\
&& +\frac{\text{mB} u (2 u-1)(1-2\sigma)}{\text{sig}-1}\,; \ea
\ba \widetilde{C}^{(X_A)}_0 &=& \frac{2 \text{q2} (1-t) t u
\big[\text{mB}^2 (\text{sig}-1)^2 +\text{q2}
(2\text{sig}-1)\big]}{\text{mB}^3 (\text{sig}-1)}
\nonumber\\
&& -\frac{\text{q2} (2 u-1)\left[\text{mB}^2
(\text{sig}-1)^2+\text{q2} (2 \text{sig}-1)\right]}{2\text{mB}^3
(\text{sig}-1)}\, ,
\nonumber\\
\widetilde{C}^{(X_A)}_1&=&\frac{u (2 u-1) \left[\text{mB}^2
(\text{sig}-1)^2+ \text{q2} (2\text{sig}-1)\right]}{\text{mB}
(\text{sig}-1)}
\nonumber\\
&&-\frac{4 (1-t) t u^2 \left[\text{mB}^2
(\text{sig}-1)^2+\text{q2} (2 \text{sig}-1)\right]}{\text{mB}
(\text{sig}-1)}\, ;
\ea
\ba C^{(Y_A)}_0 &=& -2 \text{mB} (\text{sig}-1) (1-t) t
(u-2)\,,~~~
\nonumber\\
C^{(Y_A)}_1 &=& 8 \text{mB} (\text{sig}+1) (1-t) t (u-1) u\,,
\nonumber\\
C^{(Y_A)}_2 &=& 8 \text{mB} (1-t) t u^2 \left(\frac{(3-2
\text{sig})
    u}{\text{sig}-1}+4\right)\, ;
\ea
\ba \widetilde{C}^{(Y_A)}_0 &=& \frac{2}{\text{mB}} (\text{sig}-1)
(1-t) t \big(\text{mB}^2 (\text{sig}-1)^2(u-2)
\nonumber\\
&& -\text{q2} (6 \text{sig} u+u-2)\big)+ \frac{2
\text{q2}(\text{sig}-1) \text{sig} (2 u-1)}{\text{mB}}\,,
\nonumber\\
\widetilde{C}^{(Y_A)}_1 &=& -\frac{8}{\text{mB}} (1-t) t u
\big[\text{mB}^2 (\text{sig}-1) (\text{sig} (\text{sig}(u-1)-4
u+3)+u-2)
\nonumber\\
&& + \text{q2} (\text{sig}u+\text{sig}+u-2)\big]- 4 \text{mB}
(\text{sig}-1) \text{sig}\, ,
\nonumber\\
\widetilde{C}^{(Y_A)}_2 &=& -\frac{8}{\text{mB} (\text{sig}-1)}
(1-t) t u^2 \big[\text{mB}^2 (\text{sig}-1)^2 (2
\text{sig}(u-1)-u+2)
\nonumber\\
&& +\text{q2} (2 \text{sig}+ u-2)\big] \,.
\ea

\end{document}